\documentclass[11pt,a4paper]{article}
\usepackage[dvips]{graphicx}
\begin{document}

\title{Synchronization of \\
Integrate and Fire oscillators with global coupling}
  \author{Samuele
Bottani\cite{email}\\
{\small \textit{ Laboratoire de
Physique Th\'eorique et Hautes Energies,} }\\
{\small \textit{ Universit\'{e} Pierre et Marie Curie --  Paris VI,  Universit\'{e} Denis Diderot--  Paris
VII}}\\{\small \textit{ T24-5e, 2 place Jussieu, 75251 Paris C\'edex 05,
France}} }
 
\date{}
\maketitle 
\begin{abstract}
In this article we study  the behavior of globally coupled assemblies of a large number of Integrate
and Fire oscillators with excitatory pulse-like interactions. On some simple models we show
that the additive effects of pulses on the state of Integrate and Fire oscillators  
are sufficient for the
synchronization of the relaxations of all the oscillators. This synchronization occurs in two forms
depending on the system: either the oscillators evolve ``en bloc'' at the same phase and therefore
relax together or the oscillators do not remain  in phase  but   their relaxations  occur always in
stable avalanches. We prove that synchronization can occur independently  of the convexity  or
concavity of the oscillators evolution function. Furthermore the presence of disorder, up to some
level, is not only compatible with synchronization, but  removes some possible degeneracy of
identical systems and allows  new mechanisms towards this state.
LPTHE preprint 9604
\end{abstract}

\section{Introduction}
\label{introduction}
The emergence of a large scale rhythmic activity in dynamical systems with a high number of degrees
of freedom is a widespread phenomenon occurring in different fields. In physics, macroscopic
synchronization may be found in the behavior of laser \cite{laser}, charged density waves
\cite{Fisher,StrogatzV}, networks of Josephson junctions \cite{JosephsonII,Josephson}. In Chemistry,
oscillating chemical reactions are the result of large scale synchronized activity 
\cite{chemicalI},\cite{chemicalII}. Many biological systems display also large scale
synchronization \cite{WinfreeI}. One of the most cited example is given by the south-eastern
fireflies where a large number of insects gathered on trees flash
altogether\cite{BuckI,BuckII,Hanson,ErmentroutII}. Other examples are reviewed in
\cite{StrogatzII} and include cells of the heart pacemaker, circadian neural networks, glycolytic
oscillations in yeast cells suspension, collective oscillations of pancreatic beta cells, crickets
that chirp in unison\cite{Sismondo,Walker}. Coherent oscillations are also believed to be important
in neuronal activity \cite{HopfieldI,GerstnerII}.
 
The previous systems  exhibiting large scale periodic activity are usually modeled as a large
assembly of coupled oscillators. The periodicity shown by the whole system is then the result of the 
collective synchronization of a macroscopic set of the elementary oscillators. Due to the large
diffusion of collective rhythmic behavior in nature, it is important to search and investigate all
the possible mechanisms that may lead to this phenomenon in populations of oscillators.

Most of the works related to collective synchronization in the last decade studied populations of
stable limit cycle oscillators described by ordinary differential equations continuously coupled
in time
\cite{KuramotoI,KuramotoII,Daido,DaidoII,Sakaguchi,StrogatzIII,Yamaguchi,Shiino,Matthews,Ermentrout}.
Much theoretical understanding has been obtained for such systems, as well on models where the 
phases
are the only relevant dynamical parameters 
\cite{KuramotoI,KuramotoII,Daido,DaidoII,Sakaguchi,StrogatzIII,Yamaguchi,Shiino,Hansel2,Hansel3,Hansel4}
or on models where phase and amplitude can vary\cite{Matthews,Matthews2,Ermentrout}, and for
populations of identical or almost similar oscillators. Generally, global coupling has been assumed,
i.e. each oscillator is supposed to interact with all the others. Local interactions have also been
investigated\cite{StrogatzIII,Sakaguchi,Daido} showing more complex behaviors.

These numerous studies do  not however account for the important case,
especially in biology, of episodic pulse-like interaction, where oscillating units, cells or neurons,
often communicate through the sudden firing of a pulse. Biological oscillators exchanging pulses are
currently modeled as Integrate and Fire (IF) oscillators\cite{Lapique,Glass} which are simply
described by some real valued state variable -- representing for example a membrane potential --
monotonically increasing up to a threshold. When this threshold is reached the oscillator relaxes to a
basal level by firing a pulse to the other oscillators and  a new period begins. This is the case for
example for fireflies communicating through light flashes
\cite{BuckI,BuckII,Smith,Hanson}, for crickets exchanging chirps \cite{Sismondo,Walker}, for cardiac
cells interacting with voltage pulses \cite{PeskinI} and for neurons receiving and sending synaptic
pulses.  

Large assemblies of oscillators with pulse-like coupling have been studied only  recently.
In their seminal work, Mirollo and Strogatz\cite{MirolloII} prove rigorously that a population of
identical Integrate and Fire oscillators globally coupled by exciting pulses added to the
state variables can synchronize completely for a certain kind of oscillators (convex oscillators). As
showed by Kuramoto\cite{KuramotoIII} who gives a description of such a system in terms of a
Fokker-Plank equation, the coherent collective synchronization persists when random noise is
included in the system.  Recently Corral {\sl et al.}\cite{CorralI} have generalized the Mirollo and
Strogatz model for  arbitrary evolution function of the oscillators and arbitrary response function
to pulses and established some conditions sufficient for synchronization.
When transmission delays are taken into account, Ernst {\sl et al.} 
\cite{Ernst} find that with excitatory pulses, clusters of synchronized oscillators spontaneously
form but are unstable and desynchronize after a time; partial synchronization is however achieved
with inhibitory pulses  in the form of several stable clusters of oscillators in phase. 

Other studies consider models of Integrate and Fire oscillators with the pulses smoothed before
acting on the oscillator state variable\cite{Tsodyks,Abbott,Hansel}.  Recently  Hansel {\sl et
al.}\cite{Hansel}  showed that rise and fall times can destabilize  the synchronized state of simple
IF oscillators and Abbott and van
Vreeswijk \cite{Abbott} showed that in this case the incoherent asynchronous state can be stable.
In a model with  fall times of  the coupling between the oscillators, Tsodyks {\sl et
al.}\cite{Tsodyks} 
  showed that    the complete synchronized state is unstable to
inhomogeneity in the oscillators frequencies.  Finally Gerstner
\cite{GerstnerI,GerstnerII} achieved a synthesis of results on IF oscillators models by introducing a
general model containing various versions of IF models as special cases and an  analytical approach 
from the point of view of a renewal theory \cite{Perkel,Stein,Cox}.
In the previous studies the oscillators  are typically model neurons described
as leaky integrator on a membrane potential.  This assumption determines the form of the
monotonic variation function of the state variable of the free oscillators which in this case must 
be convex.

In this article we extend a previous study \cite{BottaniI} of IF oscillators with  linear or
concave variation and with a global all to all excitatory pulse coupling directly added to the
oscillator state variables. According to the theorem of Mirollo and Strogatz\cite{MirolloII} it was
commonly believed that synchronization of pulse coupled IF oscillators could be achieved only with
convex oscillators. We show here that actually synchronization can occur independently of the shape
of the oscillators which is not therefore a  constraint for this behavior.   We present and
investigate some general mechanisms that, we think, have not been sufficiently recognized previously
and that
 lead to collective synchronization in assemblies of linear oscillators without or with quenched
disorder. These  effects,  sufficient for synchronization for linear oscillators, can also exist  for
models of leaky integrator oscillators and be combined with other mechanisms. 

The aim of this paper is not to study a particular biological or physical phenomenon in  details
but to get a better understanding of the possible mechanisms of mutual entrainment that can lead to
collective synchronization in models of IF oscillators. 
Furthermore, systems of simple linear oscillators of the kind studied in this article are also found 
in a different context than collective synchronization, which is the physics of earthquakes and
Self-Organized Criticality \cite{BakI}. This phenomenon is the spontaneous
organization of a dynamical system with a large number of degrees of freedom out of thermodynamical
equilibrium, in a critical, i.e. scale invariant, state of evolution which is attractor of the
dynamics. The building up of the long range correlations and power law behaviors characteristic of
the critical state does therefore not require the fine tuning of a control parameter (temperature,
magnetic field, etc\dots ) as for the usual critical phenomenon of  second order phase transitions.
Famous examples of dynamical systems with a high
number of degrees of freedom believed to be self-organized critical  are for example
 the sandpile model of Bak, Tang and Wiesenfeld \cite{BakI} together with several variants
\cite{DharI,ZhangI,Kadanoff}, a model of front propagation \cite{SneppenII}, evolution models for
species\cite{SneppenI}, the forest-fire model\cite{DrosselI}, etc\dots  
Self-Organized Criticality has also raised interest in Geophysics as a possible phenomenon responsible
for the scale invariant behavior of earthquakes, whose distribution of their number as a function of
their magnitude (Gutenberg-Richter distribution) is a power law. 

A classical model of earthquakes is
the Burridge-Knopoff spring-block model, where the fault between two tectonic plates is described as
a network of rigid blocks elastically connected and coupled semi-elastically and semi-frictionally to
the surfaces of the fault. Due to the relative movement of the tectonic plates, the stresses on
all the blocks increase until the stress of some block reaches an upper threshold and relaxes causing
the slipping of the block and a rearrangement of the constraints on the neighboring blocs. This can
possibly push other blocks to relax and trigger an avalanche of slippings, i.e. an earthquake. As
first noticed by Christensen \cite{ChristensenI},  the previous systems can be seen as assemblies of
pulse coupled oscillators: each block is actually an oscillator with the stress upon it acting as the
state-variable and the pulses being the sudden increment of the strain on the neighbors of the
slipping bloc. A discretized version of the Burridge-Knopoff model by
Olami, Feder and Christensen\cite{OlamiI} with linearly varying oscillators, nearest neighbors
coupling and direct action of the pulses on the state variable is believed to be self-organized
critical. It has been proposed \cite{ChristensenI,Middleton,Corral,BottaniI,HopfieldII} that the
critical behavior of this model is related to the tendency to synchronization in such systems. 
In this article we see that the globally coupled models, which are actually mean-field
versions of the Olami-Feder-Christensen model, are not critical and typically synchronize.

This paper is organized as follows:
In sections \ref{linear} and \ref{Concave} we show that because of a positive feedback 
of large groups of synchronized oscillators on smaller ones, complete synchronization of a set of
{\sl identical} oscillators is possible even in cases not taken into account by the theorem of Mirollo
and Strogatz
\cite{MirolloII}.
In section \ref{disorder}, we show how the introduction of disorder on the oscillator properties such
as the frequencies, the thresholds or the pulse strengths allows a new mechanism that can lead to
collective synchronization. Two effects act together: first, the quenched disorder makes the effective
rhythms of the oscillators all different. This brings any two oscillators to relax time
to time simultaneously. Second, oscillators that fired simultaneously possibly remain locked in a
synchronized group. 
Finally, in the last section\ref{Conclusion}, we discuss  our results focusing especially on
the effects of convexity, linearity or concavity  of the oscillators state variation function, on
additivity or not 
 of the pulses, on refractory time after a relaxation and on the possible  kinds of synchronization.

\subsection{Models}
In this article we study models of $N$ IF oscillators $O_i, i=1\dots N$ represented by a real
 state variable $E_i\in [0,E_i^c],i=1\dots N$, where the 
$E_i^c$ are the thresholds  of the oscillators. The free evolution of $O_i$ is made of two parts:
first, a charging, growth, period where the state variable $E_i$ increases
monotonically in time  as long as it is
below the threshold $E_i^c$ according to a given free evolution variation function $E_i(t)$
and, second,  a relaxation when
 the threshold is reached whereby $E_i$ is  reset to zero and a growth period
starts again. 
We  assume, as it is generally done, that the characteristic time for the relaxation
is very short compared to the  period of the free evolution so that the state variable $E_i$ of an 
oscillator that fires is instantaneously reset to zero. 

It is convenient to introduce the phases of the oscillators defined as $\phi_i\equiv t\ {\rm mod}\
\phi_i^c$ where  $\phi_i^c$ is the free period of $O_i$ ($E_i(\phi_i^c)=E_i^c$).

The coupling between biological oscillators, for instance fireflies, has been experimentally studied
by perturbing the oscillating elements by single pulses \cite{WinfreeII,Glass}.  Knowing
that fireflies interact through light flashes and that they are
believed to be describable by coupled IF oscillators \cite{BuckI,MirolloII}, the
interaction between the oscillators is studied by observing the response of the periodic flashing of a
single firefly to an artificial flash
\cite{BuckI}. Following such studies several types of couplings have been introduced in biological
models involving IF oscillators. In the situations under interest, an oscillator is coupled with
others when it relaxes and the coupling takes the form of a pulse transmitted to the others. The
consequences of the firing on the  oscillators  that have received the pulse
depend on the biological situations and on the models.

Pulses may be excitatory, i.e. incrementing the state variables and thus anticipating the firing of
the receiving oscillators, or inhibitory, i.e. decrementing the states and delaying the firing
of the receivers.

In this article we consider 
excitatory pulses:

\begin{enumerate}
\item  An oscillator receiving
a pulse has its state variable incremented by  the pulse strength. This model of coupling is 
known as phase advance model  since
the pulses push the oscillators towards their thresholds -- and possibly above -- causing a sudden
advance of the phases of the oscillators on their period of evolution.

\item The pulse strength depends on the number of oscillators that fire together and obey an
additivity principle: the pulse from the simultaneous relaxation of  oscillators is an
increasing function of the sum of all the individual pulses of the firing oscillators. For the sake
of simplicity we assume in this article direct additivity: the simultaneous firing of $n$
oscillators transmits a pulse of strength $n\delta$, with $\delta$ the  pulse strength of a single
oscillator. To account for the global coupling, $\delta$ scales  as the inverse of the system size:
$\delta=\alpha E_c/N$ with $\alpha$ a dissipation parameter.
\end{enumerate}

\section{Identical oscillators}
In this section all the oscillators are identical: $E_i(t)=E(t),\ \forall i$ and the  pulses have the
same strength. We first study the case of  linear $E(t)$ which corresponds to the limit of zero 
convexity of the model of Mirollo and Strogatz \cite{MirolloII}.

\subsection{Linear oscillators}
\label{linear}

Between two firings, the state variable increases linearly. Without loss of generality  we take simply
$E_i(t)=t\ {\rm mod}\ E_c$ so that $ 0\le E_i \le E_c=1$. 
Most studies do not consider a linear variation of the state. Indeed  
  the oscillators are commonly leaky integrators whose evolution  between
two firings is described by the differential equation:
\begin{equation}
\frac{d E_i(t)}{dt}=S_0-\gamma E_i\ \ \ \ \ , \ \ \ \ \  0\le E_i\le E_c=1,
\label{LeakyIntegrator}
\end{equation}
where $S_0$ is a constant input current and $\gamma$ describes the dissipation. The solution of
this differential equation is a convex function with the convexity controlled by the dissipation
$\gamma$.

Mirollo and Strogatz\cite{MirolloII} have rigorously  proved that with $\gamma>0$ and with constant
pulses a population of oscillators always synchronizes. From their theorem the convexity seemed to be
a necessary condition for synchronization. However, as first noticed by Christensen
\cite{ChristensenI}, a large set of oscillators with linear evolution may effectively synchronize
completely.

As we shall see, the convexity is a sufficient but
not necessary condition for synchronization. Convexity implies that the increment of the phase of an
oscillator due to a  pulse increases as the oscillator is nearer to the threshold, which has the
consequence that two oscillators effectively attract each other in the course of time.
 
We show in the following that  simply due to  the hypothesis of additivity  of pulses
there is a positive feedback effect towards synchronization in the system, which is not necessary in
the convex case for the validity of the theorem of  Mirollo and Strogatz\footnote{In \cite{MirolloII}, Mirollo and Strogatz speak of positive feedback but this
  effect plays no role in the demonstration of their theorem. This is why the
  case of linear oscillators seemed to be excluded from synchronization by
  their theorem.}. We prove
that this effect is  sufficient for synchronization even on sets of linear  and  concave oscillators.
Let us first introduce the notions of avalanche and of absorption that will be important in the
following.

\paragraph{Avalanches.}  

An avalanche of successive firings may  occur when an  oscillator reaches the  threshold:
depending on the other oscillator states the transmitted pulse may bring some other
oscillators to exceed the threshold and to fire. Possibly  the new pulses may themselves cause
further relaxations and a cascade of firings until no more pulse is sufficient  to bring
another oscillator above the threshold. In this study, we assume that the firings  and that their
transmission are very fast compared to the free evolution period   of the oscillators so that
during an avalanche the continuous drive of the oscillators is not acting.

 Avalanches are also
important for the link with the models on lattices showing Self-Organized-Criticality, which will be
discussed elsewhere \cite{BottaniII}.

\paragraph{Absorption rule and definition of synchronization.}
 
\begin{figure}[t]
\raisebox{-0.225cm}{\mbox{\includegraphics[width=6cm]{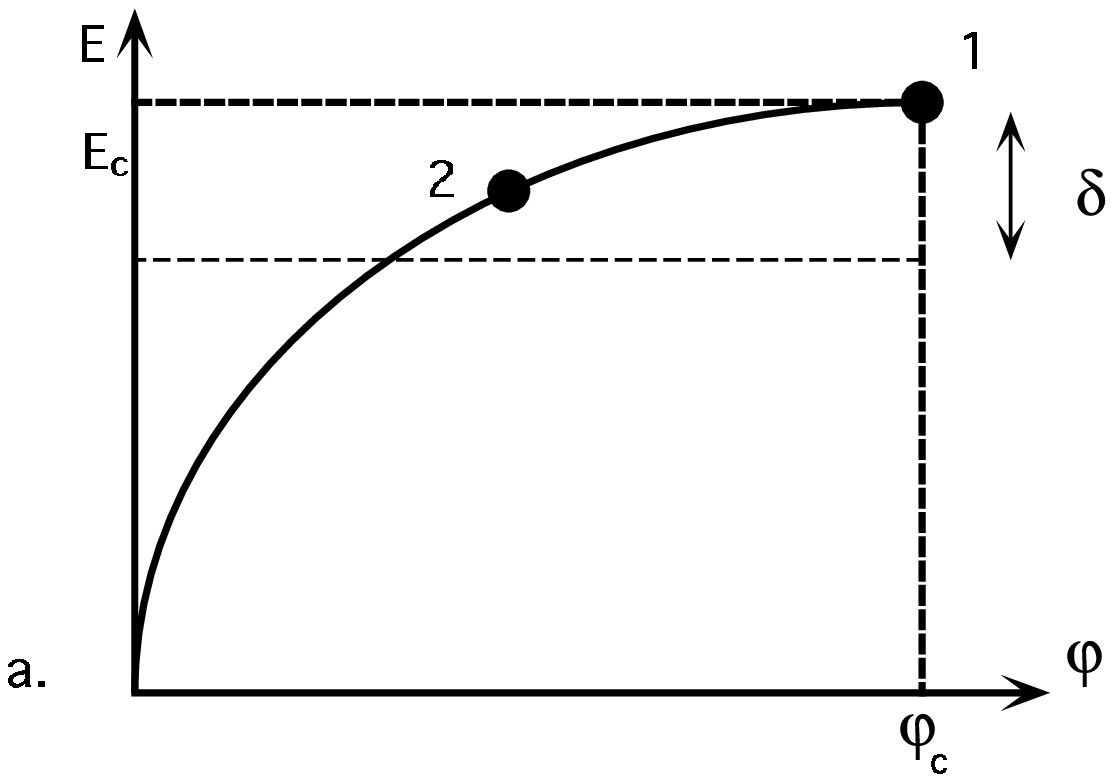}}}
\mbox{\includegraphics*[width=6cm]{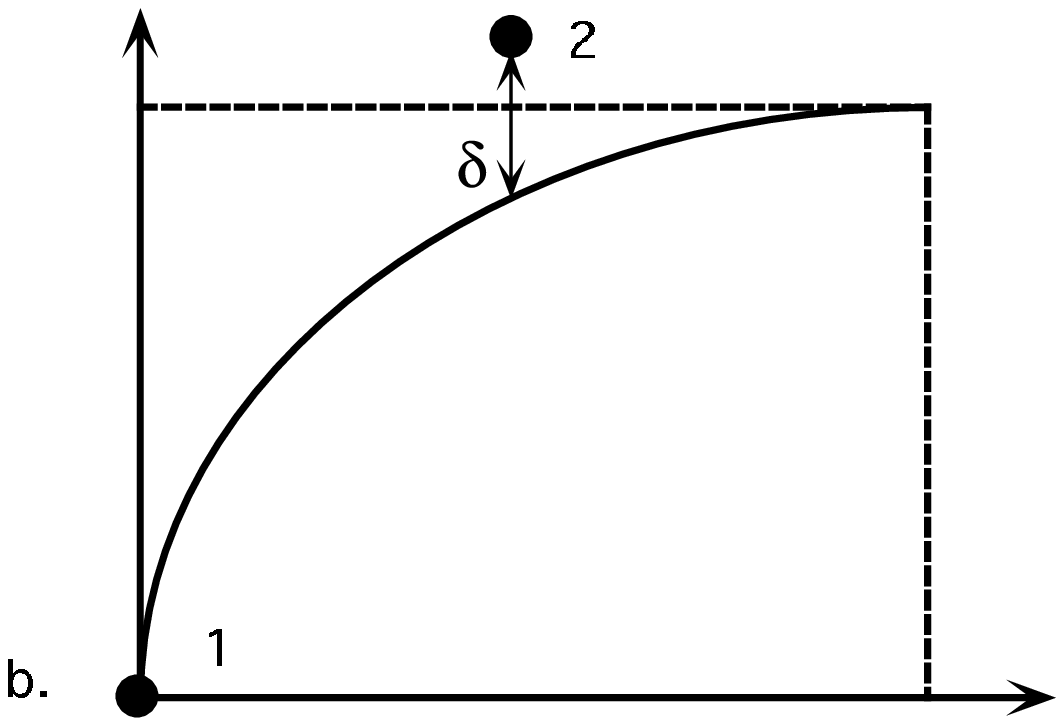}}
$$\includegraphics*[width=6cm]{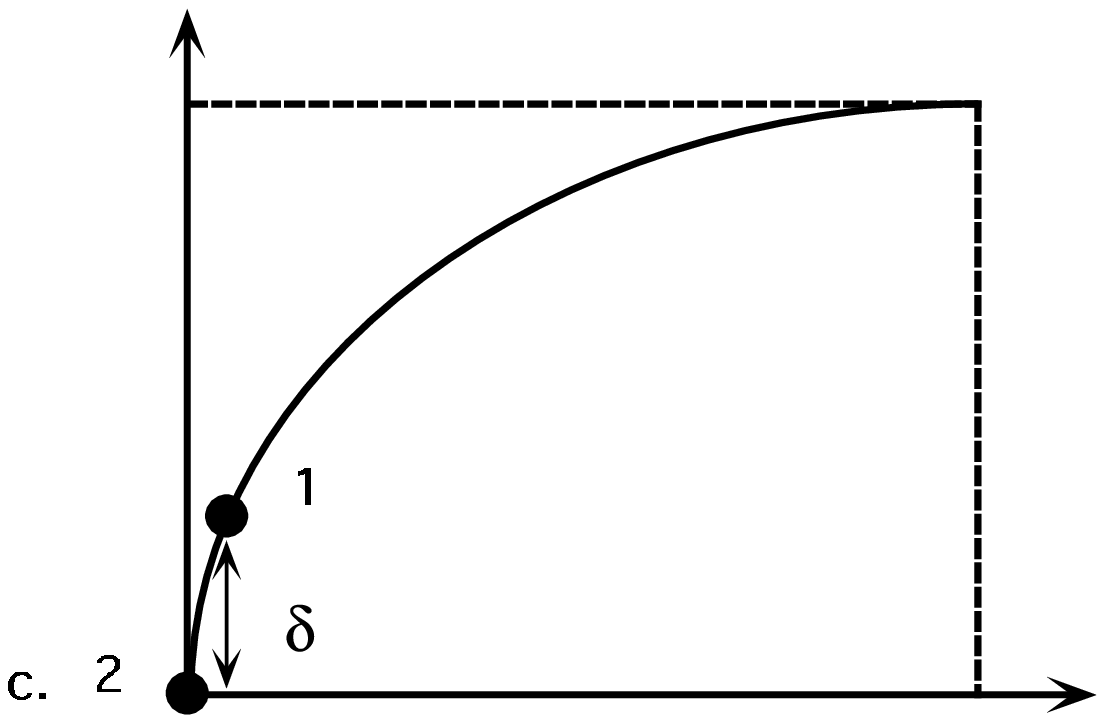}$$
\caption[Rgle d'absorption]{\renewcommand{\baselinestretch}{1}\sl Evolution without absorption. Values of the states of two identical oscillators with convex
variation. {\bf a.} The oscillator (1) is at the threshold, the oscillator (2) is below the threshold
at a distance smaller than $\delta$  which is the pulse strength of a single firing. {\bf b.} The
oscillator (1) has relaxed and the emitted pulse has pushed the oscillator (2) above the threshold and
makes it fire in avalanche. {\bf c.} Without absorption the firing of  oscillator (2) has pushed (1)
away from the origin: the oscillators remain dephased independently of the convexity. }
\label{absorption}
\end{figure}

As can be seen on fig.\ref{absorption} in the model defined up to now, oscillators can never get in
phase. A supplementary rule, that exists
also in the model of Mirollo and Strogatz, and that we call rule of absorption is necessary for
that. Since the oscillators  synchronize through the firings, we can assume,  that the oscillators
get in phase when they fire in a same avalanche.
  
\begin{figure}[p]
{\mbox{\includegraphics[width=6cm]{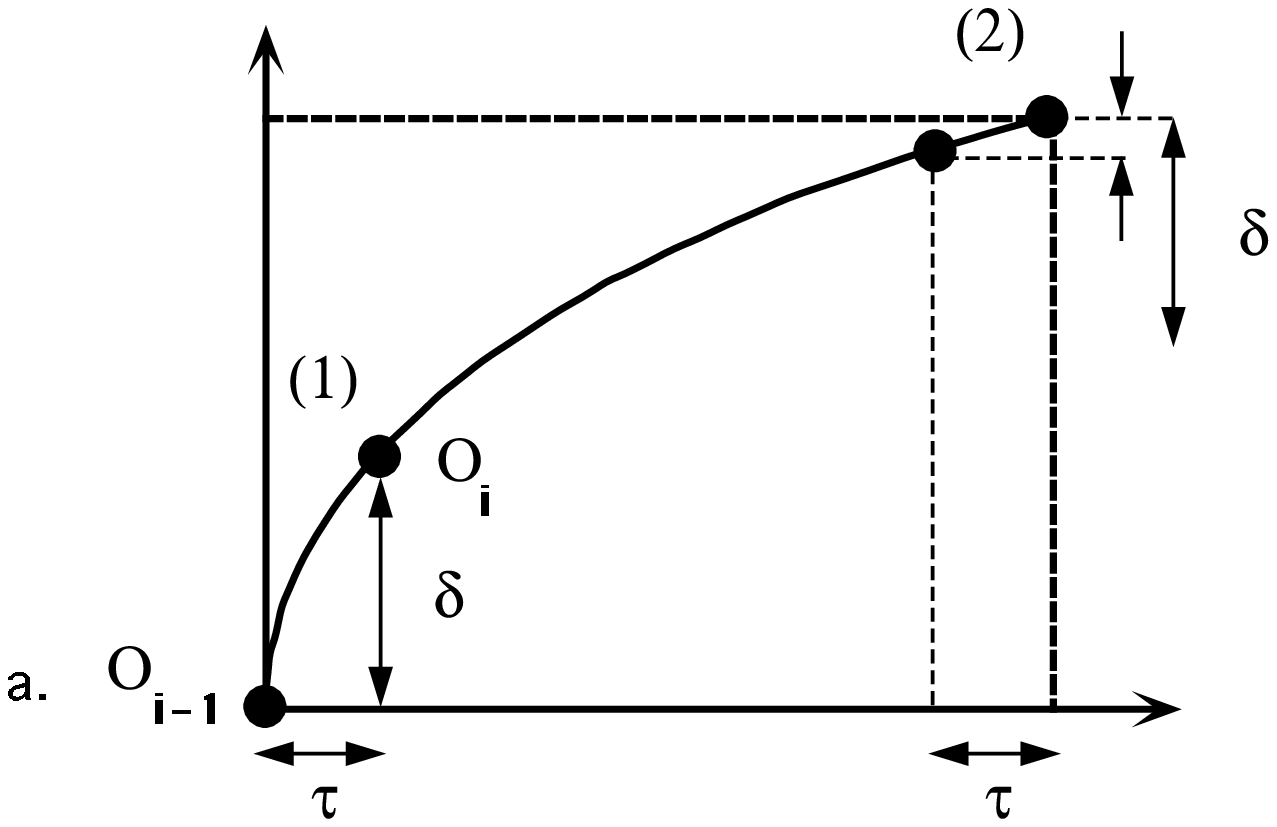}}}
\mbox{\includegraphics*[width=6cm]{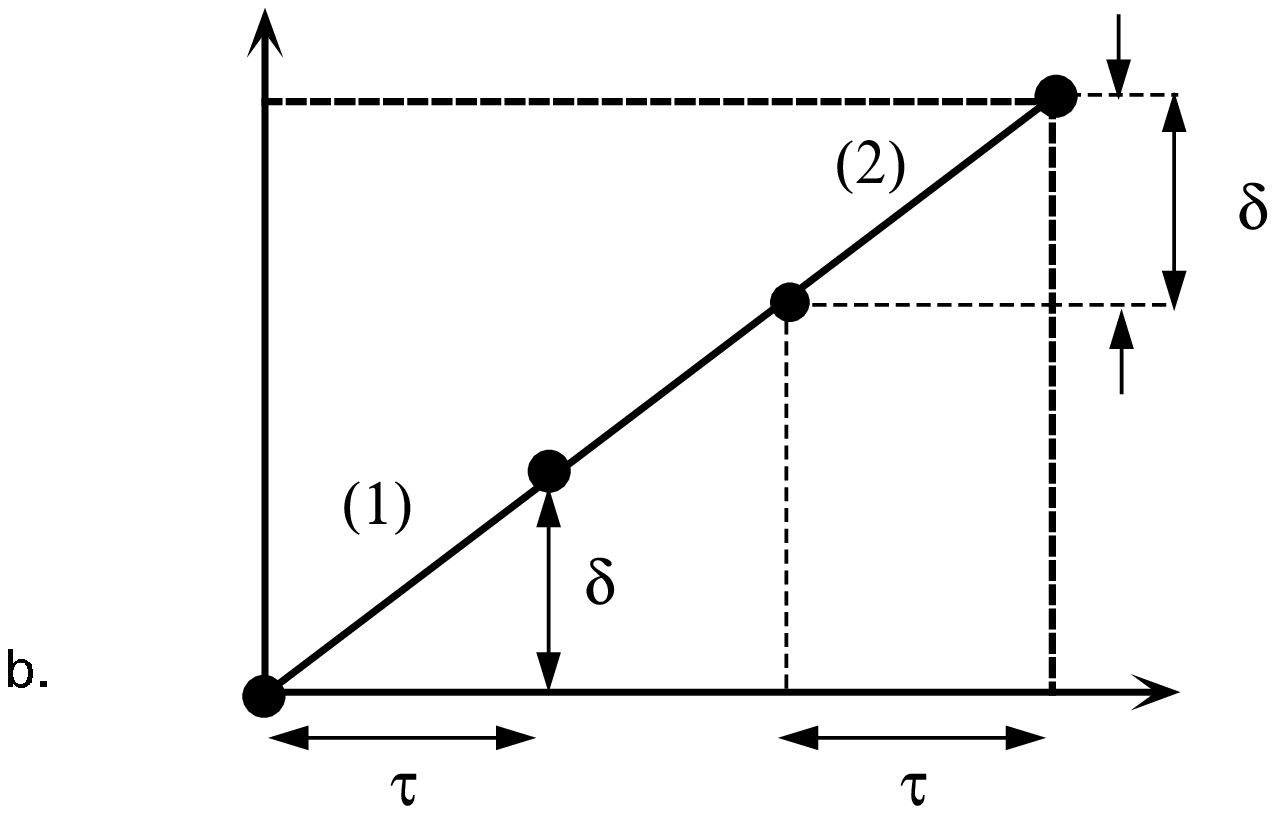}}
$$\includegraphics*[width=6cm]{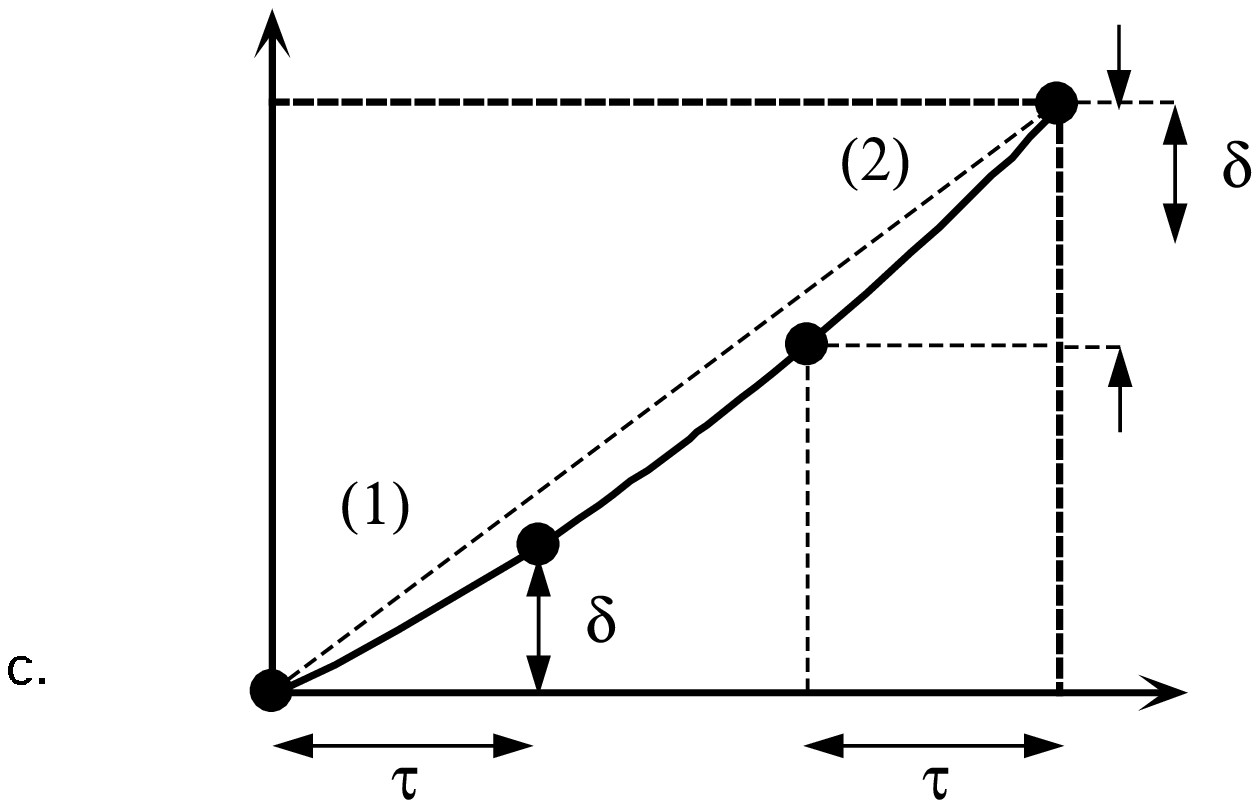}$$
\caption[Synchronisation dans des avalanches]{\renewcommand{\baselinestretch}{1}\sl{\bf a.} Synchronization
 without absorption for identical {\sl convex} oscillators.
1) Immediately after their avalanche two oscillators $O_i$ and $O_{i-1}$  have a gap between their
states
$E$ of  value
$\delta$. $\tau$ is the  gap between the phases of $O_i$ and $O_{i-1}$, which does not change
during the free evolution between firings. 2) When the most advanced oscillator is at the threshold
the gap between their phases has not changed but the gap between their state variables has decreased
due to the convexity. The second oscillator is at a distance of the threshold smaller than
$\delta$: the oscillators avalanche again together. {\bf b.} Synchronization without absorption for
identical {\sl linear} oscillators. Same as for the convex case, but due to the linearity the gap
between the states values does not change and is exactly equal to $\delta$: the oscillators  still
avalanche together. {\bf c.} Effect of concavity. The gap between the oscillators states increases
as the pair approaches the threshold.}
\label{Synchrobyavalanche}
\end{figure}

 We say that they are
  absorbed in a synchronized group of oscillators with identical phase\footnote{In \cite{MirolloII} Mirollo and Strogatz explicitly state the absorption rule
  but only for a system with numerous oscillators. Since their proof of
  synchronization is inductive with the system of two oscillators as anchor it
  is important to realize that absorption is actually necessary for
  synchronization also for a pair of oscillators.}.  
Absorption is implemented naturally by assuming that  the oscillators that relax during an avalanche
are insensitive to the further pulses in the avalanche and remain until it ends at zero  value. This
rule corresponds actually to a refractory time of the oscillators immediately after their relaxation.
Absorption is necessary for oscillators to get in phase and possibly to evolve
thereafter synchroneously with the same phase. However, it is possible to have a different definition
of synchronization in models of pulse coupled IF oscillators than evolution in phase, that does not
require the absorption rule. This is synchronization as locking into avalanches. Since we 
assume a separation of time scales between fast firings and slow continuous variation of the state
variables, locking in avalanches corresponds,  on the scale of the free oscillator period,
also to  a real synchronization of avalanches in time.
 Consider in the model without absorption two oscillators that fire in the same
avalanche as in fig.\ref{absorption}, due to the  second firing their state  are different
by  the value  $\delta$, the pulse strength of a single oscillator. When the most advanced oscillator
is back to the threshold (fig.\ref{Synchrobyavalanche}), the difference between the values of the
state variables is smaller or equal to $\delta$, in the case respectively of convex or
linear oscillators. 
In both cases the pulse from the next firing is sufficient to push the second
oscillator above or exactly at the threshold and therefore to  make it fire also: the two
oscillators are again in a same avalanche.  We see that if  two oscillators come at some time to
avalanche together they will thereafter continue to fire always together in the same avalanche also
without absorption. It is  therefore sensible to speak of synchronization also  in such a case of
locking of the firings in a same avalanche. 
We shall discuss in the rest of the paper what kind of synchronization
is possible for the different models. For systems of identical oscillators we can see on
fig.\ref{Synchrobyavalanche} that locking in avalanches
is possible for   convex or linear variation functions  but  not for concave oscillators. 
Phase synchronization is in some cases equivalent to synchronization as locking in avalanches.
Suppose that  phase synchronization  occurs in a model $A$ with  the absorption rule. 
If  in the version $B$ of the same model, but without absorption,  oscillators that are in a same
avalanche remain locked, then both models $A$ and $B$ evolve in the same way, where the same
oscillators  that are synchronized with identical phase in model $A$  are locked in an avalanche in
model $B$. Therefore if complete synchronization occurs in models $A$ then  complete
synchronization occurs also in model $B$ without absorption in the form of
locking of all the oscillators in a stable avalanche. For simplicity we choose to include 
absorption in this section on linear oscillators (here without loss of generality)  and in the
following on concave oscillators ( then necessary for synchronization).
\\
-- {\sl Proof of synchronization}
\\
Let us call configuration the set of ordered
distinct values $E_1^{(k)}< E_2^{(k)}<\dots<
E_{m_k}^{(k)}=1$ of the state variables present in the
system just before the $(k+1)$-th avalanche. To
each $E_i^{(k)}$ corresponds a group $G_i$ of $N_i^{(k)}$
oscillators at this value and
$\sum_{i=1}^{m_k}N_i^{(k)}=N$. Let call
cycle, the time necessary for all the $m_k$ groups
to avalanche exactly once. To trace the evolution of
the system, it is useful to follow, cycle after
cycle, the gaps
$s_{i,j}^{(k)}=E_i^{(k)}-E_{j}^{(k)}$ between the
values of two  groups. If one of these gaps
$s_{i,j}^{(k)}$ becomes smaller than the value
$N_{i}^{(k)}\delta$ of the pulse of the
$(i)$-th group, then the
$(j)$-th group gets absorbed by the $(j)$-th
group.
On table \ref{tablelinear} we find the main steps of the variation of the gap $s_{i,j}^{(k)}$
on a cycle beginning with $G_i$ at the threshold. Since the oscillators are identical and
linear, both groups have the same evolution as long as neither $G_i$ nor $G_j$ relax: they get the
same pulses from other relaxations with the same phase advances and between pulses their
state variables increase at the same rate.
\begin{table}

$$\begin{tabular}{|lccc|}   
\hline
&$\mathbf{G_i}$&$\mathbf{G_{j}}$&$\mathbf{s_{(i,j)}=E_i-E_{j}}$\\
\hline
{\bf a.}&$E_c$&$E_j$&$s_{(i,j)}^{(k)}$\\
{\bf b.}&0&$E_j+N_i\delta$& \\
{\bf c.}&$E_c-E_j-N_i\delta$&$E_c$& \\
{\bf d.}&$E_c-E_j+(N_j-N_i)\delta$&$0$&$s_i^{(k)}+(N_j-N_i)\delta$\\
{\bf e.}&$E_c$&$E_j+(N_i-N_j)\delta$&$s_{(i,j)}^{(k+1)}=s_{(i,j)}^{(k)}+(N_j-N_i)\delta$\\
\hline
\end{tabular}$$
\caption[Application de premier retour sur un cycle, oscillateurs linŽaires]
{\renewcommand{\baselinestretch}{1}\sl {\bf a.} Beginning of a cycle with the group $G_i$ at the threshold, the   group
$G_j$ is at a distance $s_{(i,j)}^{(k)}$. {\bf b.} Firing of $G_i$. {\bf c.} $G_j$ is at the
threshold. {\bf d.} firing of $G_j$. {\bf e.} End of the cycle $G_i$ is back at the threshold.}
\label{tablelinear}
\end{table}
 From table \ref{tablelinear} we see that the first
return map on a cycle for the gap between the oscillators is then:
\begin{equation}
s_{i,j}^{(k+1)}=s_{i,j}^{(k)}+(N_j-N_i) \delta.
\label{FirstReturnLinear}
\end{equation}
If $N_i>N_j$ the gap between the two groups decreases on each cycle. When the difference
between the states $E_i$ and $E_j$ becomes less or equal to
$N_i\delta$, then the relaxation of $G_i$ drags $G_j$ along in an avalanche. Due to the absorption,
both groups then form  a greater group with $N_i+N_j$ elements. The growth of groups is therefore due
to a positive feedback mechanism where the larger groups
attract the smaller ones. 
This effect exists only if there are in the population groups of different sizes.
 We shall now see  that as long as the number $N$ of oscillators is
sufficiently large, positive feedback always occurs until complete synchronization of the system.
If the evolution of the system begins with random initial phases for all the oscillators, all the
$E_i$ are different: there are no groups and one could naively  expect no positive feedback and no
evolution towards synchronization. However some groups are naturally formed in the first cycle of the
evolution. Indeed if two oscillators happen to be sufficiently close to  each other, i.e.   
$E_{i+1}-E_i$, the pulse from the first of them  drags the other in an avalanche and a group of two is
formed. Thereafter there are in the system single oscillators and a group of at least size two,  so
that the positive feedback mechanism can proceed.  In order to see how probable a uniform random initial 
configuration leads to the feedback effect we must therefore estimate the probability  that at least
two $E_i$ are separated by less than
$\delta$ in a set of $N$ random numbers  between zero and $E_c$. 
The probability $P(s)ds$ that two random numbers among $N$ in $[0,E_c]$ are separated by a distance
between $s$ and $s+ds$ is given in the limit $N\gg1$  by a Poissonian:
\begin{equation}
P(s)ds=\frac{N}{E_c}e^{-\frac{N}{E_c}s}ds.
\label{Poission} 
\end{equation}
The number of gaps between initial random values satisfying the condition for formation of a pair
is then:
\begin{equation}
N\int_0^{\frac{\alpha}{N}}P(s)ds=N(1-e^{-\frac{\alpha}{E_c}})\ \ \ \ \ \ ,\ \ \ \ \ \  N\gg 1.
\label{NumberPairs}
\end{equation}
Positive feedback and the absorption of oscillators into groups may take place as long as there is
at least one such a gap. It follows directly from (\ref{NumberPairs}) that this is typically the case
if
$\alpha/E_c\ge 1/N$\footnote{In \cite{ChristensenI}, Christensen concluded also that synchronization
  requires $\alpha/E_c\ge 1/N$.}. For a given level of conservation, the
number of oscillators needs only  to be large enough to ensure the onset of positive feedback.

Initial configurations  where no gap is smaller than $\delta=\alpha E_c/N$ are in principle possible.
However for large systems their occurrence is exponentially small: each gap has for $N$ large  a
probability $e^{-\alpha}$ of being greater than $\delta=\alpha/N$, so that the probability that all
the oscillators are too far apart for pair formation goes as $e^{-\alpha N}$. Therefore we may
conclude that the set of initial configurations that do not lead to absorptions is formed of
extremely improbable configurations.

To complete the proof that synchronization is the general behavior of our model, we would need to
show that the set of initial conditions for which the system evolves in a partially synchronized
configuration where  positive feedback stops acting is of almost vanishing measure.
As we have seen with the equation (\ref{FirstReturnLinear}), this can happen only when all groups are
of equal size. It is a difficult task to calculate in general the probability for a random initial
configuration to finally get stuck in such a state. 
In any case, this is a physically ill-defined problem since this probability depends critically of the
multiples of $N$: would $N$ be prime, then for every initial condition forming at least an initial
group of two the system would unavoidably synchronize completely.

Numerical simulations show that for increasing $N$ the probability for incomplete synchronization
decreases. For example with a conservation level $\alpha=0.2$   we found for $N=200,400,1000$
incomplete synchronization in respectively
$0.26,0.2$ and $0.05$ percent of the cases for $12000$ different initial configurations. For
$N=5000$ we obtained always complete  synchronization. When the
synchronization was only partial the final state of the system was always made of only two 
groups of  equal size $N/2$. For $N$ not divisible by two we found always complete
synchronization. We see that the conditions for the existence of positive feedback are almost always
fulfilled.

\begin{figure}[t]
$$\includegraphics*[width=7cm]{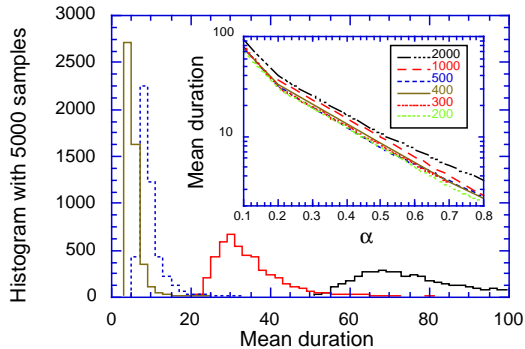}$$
\caption[DurŽes des transitoires, oscillateurs linŽaire]{\renewcommand{\baselinestretch}{1}\sl Binned distributions of the times for synchronization for a population of  $N=2000$
linear oscillators with random uniform initial phases, with conservation level
$\alpha=0.8,0.5,0.2,0.1$ and over a sample of $5000$ simulations; the time
unit is the  free period. Inset: Mean duration $T_S$ to synchronization as a function of
$\alpha$ for different population sizes, $T_S\propto \exp{-(4.3\pm 0.2)\alpha}$.}
\label{TimeSynchroLinear}
\end{figure}

We studied the time necessary
for synchronization numerically. 
Fig.  \ref{TimeSynchroLinear} shows the distribution of the durations of the transient $T_S$ until 
complete synchrony  for $N=2000$ and $\alpha=0.1, 0.2, 0.5, 0.8$.
The mean time for synchronization increases only slowly with the population size as a power
law with exponent $\sim 0.13\pm 0.01$. 
The distributions have a  flat tail towards long times
corresponding to configurations  where two groups of almost similar size remain in the system making
the positive feedback effect weak and slow to achieve the merging of the groups. 
The inset of fig.\ref{TimeSynchroLinear} shows $T_S$, calculated by cutting the tail of the
distributions, as a function of $\alpha$ for $N=200,300,400,500,1000, 2000$.
The duration of the transient $T_S$ decreases with larger conservation level and for
$\alpha\in[0.2,0.8]$ the decrease is exponential:  $T_S\propto \exp{-(4.3\pm 0.2)\alpha}$.
 Synchronization occurs then quite fast in a few free periods. The duration of the transient $T_S$
depends on the additivity of pulses. Here we assumed perfect additivity, however if the effect of the
firing of a group is not simply the sum of all the single firings but a smaller function of their
number, we expect some longer synchronization time.
We conclude that for large $N$ synchronization is possible and occurs in a finite time even for oscillators with a
linear variation which were excluded by the theorem of Mirollo and Strogatz \cite{MirolloII}.
This theorem and the older results of Peskin \cite{PeskinI} have been often erroneously  interpreted
as the necessity for synchronization of a ``leaky'' dynamics of the oscillators, which is  related
to the assumption of a convex variation function $E(t)$. 
 Let us
however stress  that the demonstration in \cite{MirolloII} for convex oscillators  proves 
complete synchronization in this case for any initial configuration apart a set of Lebesgue measure
null and is also valid  without additivity of the pulses. In the case of convex oscillators
the positive feedback mechanism is  not necessary for synchronization. Additivity of pulses and the
positive feedback mechanism that results is a further powerful mechanism which allows
synchronization under broader conditions than the effect of convexity.

 Our results with linear oscillators prove that
leaky oscillators are not necessary for the phenomenon of synchronization and that other kinds of
pulse coupled  oscillators can be considered.
As we show now, the form of the state variation function $E(t)$ is actually not even a constraint
for synchronization since this phenomenon occurs also for 
concave $E(t)$.

\subsection{Concave oscillators}
\label{Concave}

For the sake of simplicity we choose as concave function for the evolution in time of the state
variable  of the oscillators a function of the form $E(t)=f_a(t)=t^a$, with $a>1$. The effect of the
concavity on the relative state of two oscillators may be seen on fig.\ref{Synchrobyavalanche}c. Two
oscillators $O_i$ and $O_{i+1}$ with phase difference $\tau$ see the difference
$E_{i+1}(t)-E_{i}(t)$ increase as they approach  the
threshold.  Therefore with large concavity it is more difficult  that a pulse of an  oscillator 
triggers an avalanche.  However nothing forbids a group of oscillators to synchronize
if, when the first oscillator reaches the threshold, the gaps between them are smaller than the pulse
strength.

Compared with the previous case of a linearly increasing $E(t)$,  we see that now the effect of
positive feedback is  opposed by the drawing apart effect of the concavity. In a first step  we
will see that for small concavity the positive feedback effect  prevails and that synchronization
occurs. Although one would expect that for larger concavities groups would not be able to grow, we
will see in a second step  that for systems starting their evolution with an initial  random
distribution of the oscillator  phases, large concavities have the surprising consequence to favor
actually the synchronization.

\paragraph{Small concavity $a \succeq 1$}
We consider first the case of concave functions $E(t)$ that are close to the linear case. For
clarity we only sketch here   the main steps of the demonstration and refer to the appendix
\ref{appendixconcave} for  details and for the complete demonstration. We show there that for a
random initial configuration of $N$ oscillators groups begin to form and grow by positive
feedback as in the case of linear oscillators. However when only a few groups remain the positive
feedback is sufficient to reduce the phase gaps between the groups and to cause further
synchronization only if  the size differences of the groups are large enough.  The most difficult
situation for the occurrence of synchronization is when only two groups remain in the system, say 
$G_1$ and
$G_2$ with respectively
$N_1$ and $N_2$ oscillators
$(N_1>N_2)$. There is then a limit value
$\bar{c}$ of the size difference $c$ between
$G_1$   and $G_2$ so that complete synchronization occurs only if $c=N_1-N_2>\bar{c}(a,N,\alpha)$.
That is  absorption occurs only if  the size difference between the two groups is sufficiently large
so that the   positive feedback  attraction between the groups is strong and can overcomes  the
effect of concavity. Contrary to the case of linear oscillators, we see  here that two groups of
different sizes -- not only of equal sizes -- may remain apart and not synchronize. This is the
consequence of the drawing apart effect of the states by the concavity
(fig.\ref{Synchrobyavalanche}c).
Since $\bar{c}(a,N,\alpha)$ is a monotonically increasing function of $a$, for larger concavities
less final configurations  synchronize completely (for large
concavities however another effect leading to synchronization can occur, see later).

For a given $N$ there is  a
finite value $\bar{a}$ of the concavity so that $a<\bar{a} \Rightarrow \bar{c}<1$. That is, for
concavities smaller than $\bar{a}$ the system synchronizes completely unless the two last remaining
groups are of equal sizes, which is the same condition than in the linear case. 
$\bar{a}$ goes to one as $1/N$ so the corresponding range of concave functions is quite small. We
find however that synchronization occurs in practice also for much larger concavities with high
probability.

The probability $\cal{P}$ of synchronization  corresponds to the probability that the gap $c$ between
the two last groups is larger than $\bar{c}$. Unfortunately it is difficult to calculate this
probability directly. 
However we can estimate $\cal{P}$ by assuming simply a uniform distribution of $c$ in $[0,N]$.
This assumption is natural since we start the evolution with a uniform initial distribution of the
oscillator  phases. $\cal{P}$ is then the ratio of the number of favorable cases, $N-c$, over $N$,
the number of possible values of $c$. Using the value (\ref{cbar}) of $\bar{c}$  calculated in the
appendix
\ref{appendixconcave} we get:
\begin{equation}
{\cal P}=1- \frac{(1-a)}{2\alpha a}
\left(\left(1-\frac{\alpha}{2}\right)\ln\left(1-\frac{\alpha}{2}\right)+\frac{\alpha}{2}\ln\left(
\frac{\alpha}{2}\right)\right)
+o\left(\left(\frac{1-a}{a}\right)^2\right).
\label{probaconcave}
\end{equation}
Thus, the probability of synchronization is independent of the system size.
On Table \ref{ResultsConcave} we summarize the results of simulations obtained with $2000$
samples,  for $\alpha=0.5$ and several levels of concavity. We indicate
also the probabilities of synchronization expected with the assumption of uniform distribution of the
size difference of the two last groups. 
\begin{table}
$$\begin{tabular}{|lrrr|} 
\hline 
{\bf a} &$\mathbf{1.005}$&$\mathbf{1.05}$&$\mathbf{1.1}$\\ 
{\bf i.} &$99.6\pm 0.1$&$95.6\pm 0.4$&$90.6\pm 0.6$\\
{\bf ii.} &$99.4$&$94.4$&$89.1$\\
\hline
\end{tabular}$$
\caption[ProbabilitŽs de synchonisation, oscillateurs concaves]{\renewcommand{\baselinestretch}{1}\sl{\bf i.} Probabilities of complete synchronization  with the statistical
error obtained with
$2000$ samples for $\alpha=0.5$ for ``small'' concavities $a=1.005,1.05,1.1$. The probabilities 
obtained with 
$N=2000,1000,500,400,300,200$ are identical within the error. {\bf ii.} Estimated probabilities of
complete synchronization with the assumption of uniform distribution of the size difference between
the two last  groups in the system.}
\label{ResultsConcave}
\end{table}
Within the statistical error the probabilities of
synchronization are independent of
$N$  and correspond to  the expectations.  

\begin{figure}[t]
$$\includegraphics*[width=7cm]{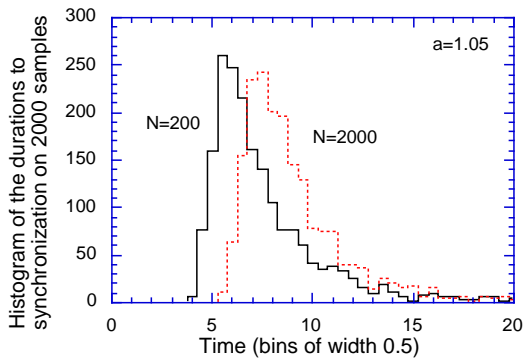}$$
\caption[DurŽes des transitoires, oscillateurs concave]{\renewcommand{\baselinestretch}{1}\sl Binned distributions of the duration $T_S$  of the transient until complete synchronization
in a population of $N=2000$ identical concave oscillators with concavity $a=1.05$ for $2000$ samples
of uniformly distributed random phases.}
\label{DurationConcav}
\end{figure}

For small concavities $a=1.005, 1.05, 1.1$ we found that the duration $T_S$ of the transient
until synchronization does not depend on the value of the concavity. In fig.\ref{DurationConcav} we
report  the distributions of $T_S$  for $a=1.05$. It can be seen that
typically synchronization occurs in a few free periods.  
 Furthermore 
$T_S$ increases only slightly with the population size as a power law with a small exponent:
$T_S\propto N^{0,09\pm 0.01}$ for $N=200$ to
$2000$ and $\alpha=0.5$. Large populations synchronize therefore quite fast as in the linear case.

From what
precedes we would expect that synchronization is impossible for large concavities. Without entering
into details we shall now see  that assuming a natural uniform initial distribution of the oscillators
phases (and not of the states $E_i$) there is for large concavities a cross-over in the behavior of
the system towards easier synchronization. 

\paragraph{Large concavities}

\begin{figure}[t]
\mbox{\includegraphics[width=6cm]{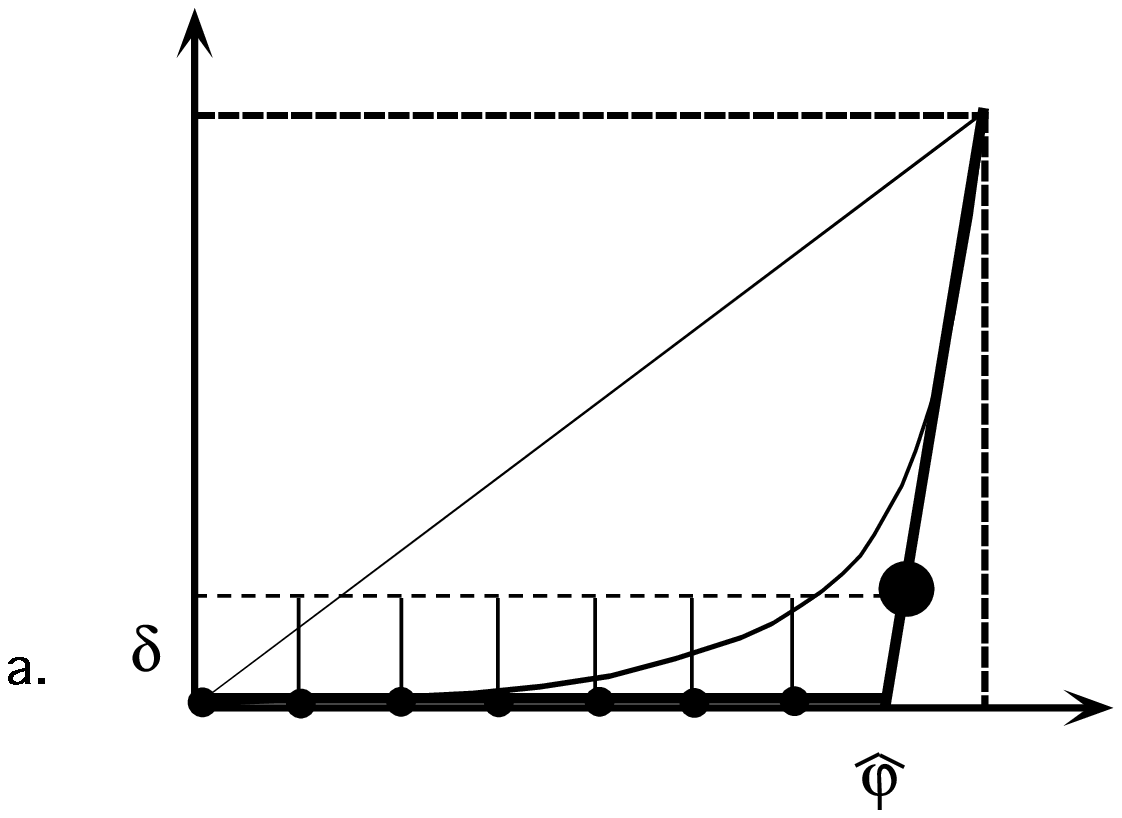}}\mbox{\includegraphics*[width=6cm]{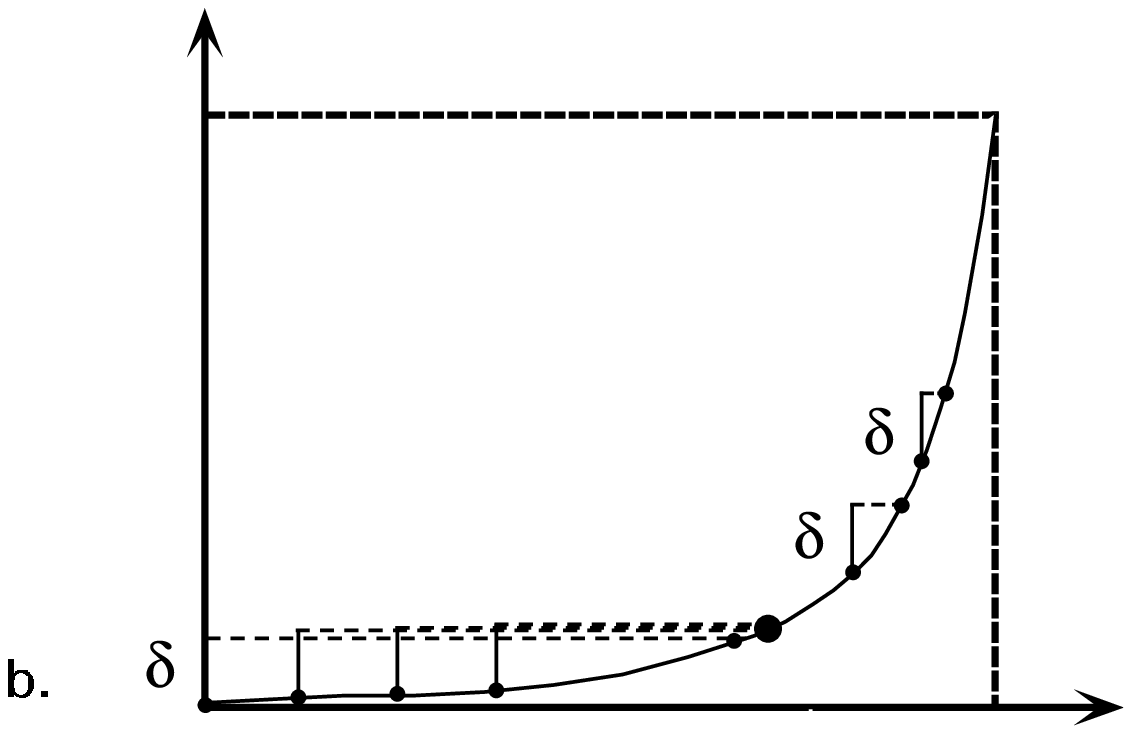}}
\caption[Grandes ConcavitŽs]{\renewcommand{\baselinestretch}{1}\sl {\bf a.}``Extremal'' model with very large concavity. The state value evolution function is
flat up to some phase $\hat{\phi}$ where it abruptly monotonically increases up to the threshold. All
the oscillators with initial phase smaller than $\hat{\phi}$  have the same state value. They
synchronize at the same phase value as soon as they  receive any pulse
$\delta$. {\bf b.}  For a large concavity $a$ but with a smooth state value evolution function
$E(t)=f_a(t)$, the oscillator with small initial phases have also very close initial states. A pulse
$\delta$ brings them at almost the same phase.}
\label{bigconcavity}
\end{figure}

Let us first illustrate the mechanism at work on an extremely simplified model shown on
fig.\ref{bigconcavity}a. where we replace the concave function by the union of its tangent segments at
both extremities. 
That is the free evolution function of the oscillators is now:
\begin{equation}
E(\phi)=\left\{\begin{array}{ll}
0&, \phi  \in [0,\frac{1}{a-1}]\\
(a-1) \phi+1&, \phi \in [\frac{1}{a-1},1]
\end{array}
\right.
\end{equation}
with $a>1$. 
 All the oscillators with initial phases in
$[0,1/(a-1)]$  have the same initial state value $E_i=0$ due to the special form of the evolution
function. It is clear from fig.\ref{bigconcavity}a
that all these oscillators get in phase and synchronize  as soon as  the first pulse of the evolution
occurs. It is then possible to show that the large synchronized group that is thus formed absorbs
thereafter the  oscillators that where initially in $[1/(a-1),1]$ and the system synchronizes
completely.

For smoother $E(t)$ the same mechanism occurs  (see
fig.\ref{bigconcavity}b). For
small phases $\phi\rightarrow 0$ the slope of the evolution function is small and thus the $E(\phi)$
are closer to each other than for larger phases where the
slope is steeper. 
 If the initial packing of the
states  subsists  until one of the closely packed oscillators is at the threshold a large avalanche
occurs and thus  the synchronization of many oscillators. It is not  obvious
that the states remain close to each other.
 Indeed during the free evolution of the system (between
firings) the gaps between the phases do not change  but  the states get farther apart from each other
due to the concavity. 
On the opposite, during firings     the gaps 
between the states remain constant since all the states are incremented the same way (whereby the
phase gaps get smaller). Let the oscillators be numbered by increasing order of their initial phases
$\phi_{i+1}^{(0)}>\phi_i^{(0)}$. The evolution of $O_i$ towards the threshold is caused as
well by free evolution between avalanches as by phase advances due to pulses.  Before reaching the
threshold, an oscillator $O_i$ receives
$N-i$ pulses from the oscillators with larger initial phases.  For small initial phases
$(i\rightarrow 1)$ the oscillators $O_i$ and
$O_{i-1}$ receive many pulses and their evolution towards the threshold is for a large part
due to the  phase advances from pulses.  Possibly there is sufficient evolution due to pulses so
that the state gaps do not increase enough  due to the free evolution to prevent a large avalanche
of the initially closely packed oscillators.

\begin{table}
$$\begin{tabular}{|crrrc|} 
\hline
{\bf a}&$\mathbf{500}$&$\mathbf{1000}$&$\mathbf{2000}$&{\bf estimate} $\mathbf{a\succeq 1}$\\
$\mathbf{1.55}$& $68\pm1.0$&$83\pm0.8$&$95\pm0.5$&$50$\\
$\mathbf{2.}$&$93\pm0.5$&$99.9\pm0.1$&$100$&$25$\\
\hline
\end{tabular}$$
\caption[ProbabilitŽs de synchronisation, grandes concavitŽs]{\renewcommand{\baselinestretch}{1}\sl {\bf i.}
 Probabilities of complete synchronization with the statistical error for ``large''
concavities
$a=1.55$ and $a=2$ with $N=500,1000,2000$ and an a uniform initial distribution of the phases. The
last column right shows the probabilities expected as for small concavities.}
\label{ResultsConcaveII}
\end{table}
On Table (\ref{ResultsConcaveII}) we see that for concavities $a=1.55$ and $a=2$
synchronization already occurs with a larger probability that expected with  the estimate  from
small concavities. As expected the probability for synchronization increases with $a$
for a given population size $N$. We see also that the probability increases with $N$. This is the
consequence that with a uniform distribution of phases the oscillators
are  at the beginning denser  for larger populations  in the flat section of $E(\phi)$
and larger synchronized groups form at the beginning of the evolution thus enhancing the positive
feedback.
Large concavities favor synchronization only for a uniform distribution of the initial phases. 
Indeed if, instead of the phases, the states
$E_i$ of the oscillators were initially uniformly distributed in
$[0,1]$, there would be, per definition, no clustering and the oscillators would stay apart, 
as it is easy to verify looking at
fig.\ref{bigconcavity}a. An
initial uniform distribution of the  phases is however a natural assumption for the beginning of the
evolution.

Finally, the main conclusion  of this section is  that surprisingly   the form of
the oscillator state variation function $E(\phi)$ is not actually relevant for synchronization that
occurs with high probability for functions $E(\phi)$ which are convex, linear  and even  concave
provided the phases are randomly distributed initially.  The usual
interpretation of ``leakiness'' (implying convexity) as a requirement for synchronization must
therefore  be revised. 
Up to now we have considered only
identical oscillators.  In the following sections we show that in populations of oscillators with
different randomly distributed characteristics, synchronization occurs also in a different way that
what we have seen up to now.    

\section{Systems with quenched disorder}
\label{disorder}
We shall see that with quenched disorder, synchronization is the combined consequence of several
causes. For the sake of simplicity we show how synchronization occurs in the  
cases of oscillators with respectively different free  frequencies then    different
amplitudes and finally of oscillators with different  frequencies and
amplitudes.  The mechanisms at work are the same for the different kinds of disorder although some
important peculiarities depend on the models. In brief, these mechanisms and the main steps of the
demonstrations are the following.

 We write  the first return map for the phase of a given oscillator $O_j$ on a cycle beginning
and finishing when another given  oscillator $O_i$ is at the threshold. During such a cycle 
all the oscillators of the system fire once. The first return map  shows that due to
the quenched disorder $O_i$ and $O_j$ inevitably fire at some time, after some cycles,  in a
same avalanche independently of the initial values of their states.
After their relaxation, oscillators that have fired together are  at the origin  and in phase  such
that $E_i=E_j=0$ (assuming a refractory
time).
However,  contrary to the case of identical oscillators, the fact that they have simultaneously
relaxed together  does not imply  that they  will forever continue to fire 
together. Indeed different intrinsic rhythms or different
 responses to pulses (see later) dephase the oscillators that were in phase.
However, it is physically clear that for oscillators with sufficiently close characteristics
(frequency, threshold, shape etc\dots), the disorder cannot destabilize a group of oscillators that
have fired once together. More precisely, it is possible to state stability conditions that have to
be fulfilled by any group of oscillators that have fired together in order to remain synchronized.
Since any two oscillators necessarily fire at some time simultaneously, all the possible groups
fulfilling stability conditions are formed during the evolution.
If the stability conditions are fulfilled by the whole oscillator population larger, groups
progressively form up to complete synchronization independently of the initial values of the
$E_i$. The probability for complete synchronization is therefore the probability that a random
sample of oscillators  fulfills the  stability conditions on the whole  system.

\subsection{Distribution of frequencies}
\label{frequencies}
In this section we consider models of linear IF oscillators with a spread of intrinsic frequencies.
Since we shall not allow adaptation\footnote{By adaptation of the frequencies we mean as in \cite{ErmentroutII} a
  ``learning'' mechanism, where the oscillators are allowed to modify their
  intrinsic frequencies in order to match with an exterior periodic stimulus.}
of the free frequencies, two oscillators that fire once simultaneously do not subsequently reach the threshold at
the same time and in general do not fire simultaneously again. For a system with a spread of the intrinsic frequencies we shall
therefore consider synchronization of oscillators as relaxation in the same avalanche, which corresponds to temporal synchronization
in the limit of very short characteristic time for the transmission of the pulses  compared to the period of free evolution (see also
\ref{linear}).
Let in our model all the oscillators be identical apart from their free periods $\phi_i^c$ which
are uniformly randomly distributed in an interval $[\phi_{min}^c,\phi_{max}^c]$. Without loss of
generality we take their common slope equal to one so that each oscillator has a threshold
$E_i^c=\phi_i^c$.   The pulse strengths of all the oscillators are supposed to be identical and equal
to
$\delta=a\alpha/N$ with $a=(\phi_{min}^c+\phi_{max}^c)/2$ the center of the distribution interval
of the periods.

\begin{table}
$$\begin{tabular}{|lccl|}
\hline
&$\mathbf{\phi_i}$&$\mathbf{\phi_j}$&\\
\hline
{\bf a.}&$\phi_i^c$  &$\phi_j^{(k)}$&\\
{\bf b.}&$0$&$\phi_j^{(k)}+\delta$&\\
{\bf c.}&${\Delta_1}$&$\phi_j^{(k)}+{\delta+\Delta_1}$&\\
{\bf d.}&$\phi_j^c-\phi_j^{(k)}-{\delta}$&$\phi_j^c$&\\
{\bf e.}&$\phi_j^c-\phi_j^{(k)}$&$0$&\\
{\bf f.}&$\phi_j^c-\phi_j^{(k)}+\Delta_2$&${\Delta_2}$&\\
{\bf g.}&$\phi_i^c$&$\phi_j^{(k+1)}=\phi_j^{(k)}+(\phi_i^c-\phi_j^c)$&\\
\hline
\end{tabular}$$
\caption[Application de premier retour, dŽsordre de frŽquences]{\renewcommand{\baselinestretch}{1}\sl {\bf a.} Oscillator $O_i$ is at the threshold; {\bf  b.} Firing of $O_i$ assuming that
$O_j$ is not pushed above the threshold; {\bf c.} Effect of the sum $\Delta_1$ of all the pulses from
other oscillators of the system between the firings of
$O_i$ and the one of $O_j$; {\bf d.} $O_j$ at the threshold; {\bf e.} Firing of $O_j$; {\bf f.}
Effect of  the sum $\Delta_2$ of  the pulses between the firings of $O_j$
and  $O_i$ back at the threshold; {\bf g.} $O_i$ back at the threshold.}
\label{TableFrequencies}
\end{table}
We follow the steps of the demonstration of synchronization outlined before.  
From Table~\ref{TableFrequencies} we see that the first
return map of the phase of $O_j$ on  a cycle is
\begin{equation}
\phi_j^{(k+1)}=\phi_j^{(k)}+(\phi_i^c-\phi_j^c).
\label{FirstReturndisorder}
\end{equation}

Since the periods are random parameters, $(\phi_i^c-\phi_j^c)$ is typically a non
zero constant. If this difference is positive $O_j$ comes closer to its  threshold $E_j^c$ at each
cycle beginning with  $O_i$ at  $E_i^c$ . Therefore after some
cycles  the firing of $O_i$  drags $O_j$ along in an avalanche.  If 
$(\phi_j^c-\phi_i^c)<0$   we are in the previous situation  by interchanging 
$O_j$ and  $O_i$. In any case the conclusion is the same: at some time the
two oscillators fire in a same avalanche. 

\begin{figure}[t]
$$\includegraphics*[width=7cm]{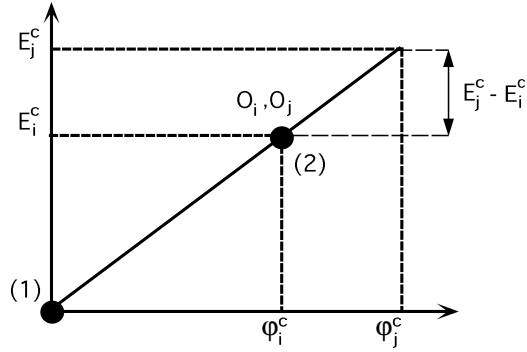}$$
\caption[Distribution de frŽquences]{\renewcommand{\baselinestretch}{1}\sl Two oscillators $O_i$ and $O_j$ with different periods $\phi_i^c$ and 
$\phi_j^c, (\phi_i^c<\phi_j^c)$ and identical slope. In $(1)$ the two
oscillators have just relaxed in an avalanche and are in phase at the origin; both oscillators evolve thereafter in phase. In $(2)$
 the oscillator $O_i$ with the highest frequency is at his threshold $E_i^c$. 
$O_j$ is at a distance $E_j^c-E_i^c$ below its
threshold.}
\label{FigureFrequency}
\end{figure}

Just after their relaxation in the same avalanche, the states and phases of $O_i$ and $O_j$ are both
at zero.  Since the oscillators have the same slopes the pulses from the rest of the system increment
the phases of $O_i$ and $O_j$ with the same value and both oscillators evolve therefore in parallel
with
$E_i=E_j$ until the oscillator with the highest frequency, say $O_i$, reaches first its threshold
$E_i^c$. When $O_i$ fires, $E_j=E_i^c$ and is therefore below its threshold $E_j^c$ (fig.
\ref{FigureFrequency}). 
Both oscillators remain synchronized only if the pulse from $O_i$ is
sufficient to push $O_j$ above its threshold, so that the stability condition for a pair
oscillators is:
$E^c_i+\delta\ge E_j^c$ (equivalently since the slope is equal to one, $\phi^c_i+\delta\ge
\phi_j^c$ ). 

\begin{figure}[t]
$$\includegraphics*[width=7cm]{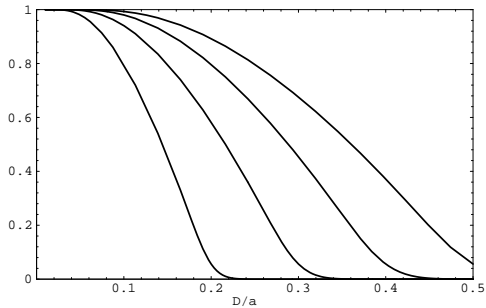}$$
\caption[ProbabilitŽ de synchronisation avec une distribution de frŽquences]{\renewcommand{\baselinestretch}{1}\sl Probability for a configuration allowing complete synchronization in a system of
oscillators with a uniform random distribution of intrinsic frequencies. The probability depends only
on the ratio $D/a$ where $D=\phi_{max}-\phi_{min}$ and $a=(\phi_{max}+\phi_{min})/2$  and on the
conservation parameter $\alpha$. From left to right
$\alpha=0.2,0.3,0.4,0.5$.}
\label{GraphprobaFreq}
\end{figure}

More generally, for larger groups, suppose that $n$ oscillators $O_i, i=1\dots N$ with
$\phi_{i+1}^c>\phi_i^c$ just fired together in an avalanche so that $\phi_i=0, E_i=0\ \forall i$.  The
oscillator
$O_1$ with the shortest period (threshold) is the first to reach again its threshold. It triggers an
avalanche involving the $n-1$ other oscillators if
\begin{equation}
\phi_{i+1}^c-\phi_1^c\le i\delta\ \ \ ,\forall i=2\dots n.
\label{conditiongroup}
\end{equation}
This condition comes from the fact that the $i+1$-th oscillator receives in the avalanche a total
pulse $i\delta$.
A random configuration of  frequencies may allow complete synchronization if the
inequalities (\ref{conditiongroup}) are fulfilled for all the oscillators of the system $(n=N)$. 
The probability ${\cal P}_N$ for a system with a random uniform distribution of $N-2$
periods in $[\phi^c_{min},\phi^c_{max}]$ to allow complete synchronization
 is the product of
the  probabilities for each gap $s_i\equiv (\phi^c_{i+1}-\phi^c_1)$ to be smaller than  $i\delta$.
Since $s_{i+1}>s_i$  we get after a change of variables:
\begin{eqnarray}
{\cal P}_N&=&\rho^N\int_0^\delta ds_1\int_{s_1}^{2\delta}ds_2\dots\int_{s_{N-1}}^{N\delta}ds_N 
e^{-\rho s_N}\\
&=&1-e^{-\rho\delta}-\rho\delta e^{-2\rho\delta}-\sum_{j=2}^{N-1}\frac{(j+1)^{j-1}}{j!}
(\rho\delta e^{-\rho\delta})^je^{-\rho\delta},
\label{probafrequency}
\end{eqnarray}
where
$\delta=(\frac{\phi^c_{max}+\phi^c_{min}}{2})\frac{\alpha}{N}$ is the pulse strength and 
$\rho={N\over \phi^c_{max}-\phi^c_{min}}$  is the uniform density of the intrinsic periods. This
probability depends only on the ratio
$D/a$ of the width $D$ of the distribution $(D=\phi_{\rm max}^c-\phi_{\rm min}^c)$ and on the  center
$a=(\phi_{\rm max}^c+\phi_{\rm min}^c)/2$ of the distribution through
$\rho\delta=D/a$. The probability (\ref{probafrequency}) is
plotted in fig.\ref{GraphprobaFreq} for $\alpha=0.2,0.3,0.4,0.5$ and
$N=300$. We see that for a finite width $D$ a large fraction of  the initial samples of randomly
distributed periods allows complete synchronization, typically for $D/a<0.1$ and $\alpha>0.2$
synchronization occurs in  more than
$95\%$ of the cases. 
Nevertheless, after a flat section at small widths, $\cal{P}$ decreases rapidly
with increasing $D/a$.  
Therefore, although complete synchronization is possible with very high probability for small $D/a$,
the range of disorder on the frequencies compatible with this behavior is
limited.  
In the region of high
synchronization probability we find that $\cal{P}$ is unaffected by the population size when $N$ is
large (typically $\succeq 100$) since in this limit only the tails of the distributions at
large $D/a$  actually  depends on $N$.
 We have studied the duration of the transient $T_S$ until synchronization  numerically on
simulations (see fig.~\ref{DurationFreqandThresh} inset). Up to $N=500$ we found that
$T_S$ increases linearly with $N$ with a small slope. For instance with $D=0.2, a=1, \alpha=0.5$
we have
$T_S\simeq 19+0.06 N$ (fig.\ref{DurationFreqandThresh} inset). Since the divergence of $T_S$ with
$N$  is only linear, synchronization occurs in a physically reasonable time.

In the previous model the system synchronizes at the  frequency of the fastest oscillator. This is a
direct consequence of the absorption rule that sets at the origin  all the oscillators that
participate in an avalanche. It is therefore interesting to study the same model but without the
absorption rule. Let us remind that for identical linear oscillators synchronization, as locking
in avalanches, was still possible without absorption. \label{SansAbsorption}
For the model with a spread of frequencies the first return map (\ref{FirstReturndisorder}) is valid
also without absorption. Let us take two oscillators $O_i$ and $O_j$ with  $\phi_j^c<\phi_i^c$. At
some time $O_i$ drags $O_j$ in an avalanche: $O_i$ fires and relaxes to zero and, without
absorption, is immediately incremented to $E_i=\delta$ by the following firing of
$O_j$. Therefore after the avalanche the oscillator $O_i$  is more advanced in
phase and  the oscillator with the highest frequency $O_j$ 
does not necessarily reach its threshold first, contrary to the case with absorption. 

It is easy to verify that $O_i$ is the first of the two oscillators to reach its threshold $E_j^c$ if
$\phi_i^c-\phi_j^c<\delta$  i.e. $E_i^c-E_j^c<\delta$. In this case the firing of $O_i$
automatically drags $O_j$ again in an avalanche since we have $E_i-E_j=\delta$ and thus
$E_i^c-E_j^c <\delta\Rightarrow E_j^c-E_j=E_j^c-E_i^c+\delta<\delta$. The two of oscillators are
therefore locked in an avalanche and form a stable group that fires with the longest period
$\phi_j^c$ of the two. 
If $\phi_i^c-\phi_j^c>\delta$ then $O_j$ is first at its threshold and fires before $O_i$. Although 
it is possible that $O_i$ and $O_j$ avalanche again together this time,  the two oscillators can not
remain locked in an avalanche further. Indeed, $E_j-E_i=\delta$ since $O_j$ fired first and when 
$O_j$ is back at $E_j^c$  we have $E_i^c-E_i=E_i^c-E_j^c+\delta>\delta$ so that $O_i$ does not
avalanche with $O_j$. 

We see that without absorption synchronization of two oscillators is still possible  but at the
lowest frequency. This result can be straightforwardly generalized to $N$ oscillators following the
same procedure as in the case with absorption. We find actually that the locking conditions for the
whole system are a set of inequalities equivalent to (\ref{conditiongroup}) so  that the
probability of complete synchronization for a uniform distribution of $\phi_i$  in
$[\phi_{min},\phi_{max}]$  is given by the same expression as (\ref{probafrequency}).

Let us just mention that the fact that  synchronization  occurs also without absorption is
important for the behavior  of some lattice models of oscillators displaying SOC
\cite{Middleton,BottaniII}. In these models the oscillators are locally coupled by pulses without
an absorption rule. As first shown by Middleton and Tang on the Olami-Feder-Christensen
model\cite{Middleton}, depending on the number of nearest neighbors,  oscillators have different
effective frequencies. From what precedes, we would expect some synchronization in the system and
indeed,  a tendency towards synchronization is observed also on the lattice. Complete
synchronization does not occur, but there is partial synchronization at  all scales \cite{BottaniII}.

We see finally that in a simple model of IF oscillators with a spread
of the free frequencies synchronization can occur in the form of locked avalanches with or without
the absorption rule, i.e. a refractory time. However the presence or not of the absorption rule
changes drastically the nature of the synchronized avalanches which are respectively triggered by the
oscillator with highest and shortest free frequency. This sensibility to the absorption rule together
with a probability of synchronization strongly dependent, above some value, on the distribution width
indicates that, apart in some  limits, in a real situation synchronization is restricted  by a
disorder on the frequencies.

In a model with pulses with a finite fall time
Tsodyks {\sl et al.}\cite{Tsodyks}  showed that  synchronization is unstable. However, our results
 show that synchronization is not incompatible in principle with a disorder in frequencies in
pulse coupled oscillators models in the limit of short instantaneous pulses and when the notion of
synchronization in avalanches is valid.

\subsection{Oscillators with different amplitudes}
\label{amplitudes}
\begin{figure}[t]
$$\includegraphics*[width=7cm]{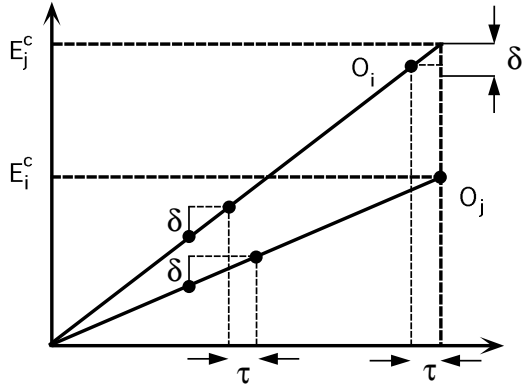}$$
\caption[Oscillateurs avec diffŽrentes amplitudes]{\renewcommand{\baselinestretch}{1}\sl Oscillator with different amplitudes $E_i^c$ and equal frequency. With $\phi^c=1$ the
oscillators have different slopes $a_i=E_i^c$. 
A pulse of strength $\delta$ dephases two oscillators that avalanched together and were in phase at
the origin. The oscillator $O_j$ with lowest slope gets the largest phase advance ($\delta/a_j$) and
reaches  the threshold  before  $O_i$. If the state $E_i$ of $O_i$ is at a distance to its
threshold $E_i^c$ smaller than the value $\delta$ of the pulse of $O_j$ then the two
oscillators stay synchronized.}
\label{distribampl}
\end{figure}

In this section we keep the frequencies of the oscillators equal (the period is $\phi_i^c=1,\forall
i$) and let the thresholds have different values (fig.\ref{distribampl}).
 Each oscillator $O_i$ is
then characterized  by a
 threshold $E_i^c$ and  has a  slope $a_i=E_i^c$. By disorder on the amplitudes we mean therefore
disorder on the thresholds with  related distribution of the slopes.
We
keep the pulse equal for all the oscillators:  $\delta=\alpha/N$. 
Since all the oscillators have the same free period, synchronization  in the sense of
variation in phase of all the  oscillators and simultaneous relaxations is possible in this
model. 
We follow the same steps as previously.
Since the mechanisms at work are similar than in the model with a distribution of frequencies we
leave the details of the discussion in appendix (\ref{appendixamplitudes}). Since  the frequencies
are now equal and the slopes and thresholds are different, the main differences with the preceding
section are in the reasons why simultaneous firings occurs and groups form.
Here also the phase gap between any two oscillators changes monotonically after each cycle, so that
any two oscillators avalanche at some time together.  The change in the phase gaps between the oscillators that finally cause the
simultaneous firings 
 has for origin also here  the different  rhythms of firings of the oscillators. But contrary
to the previous model where the different rhythms were intrinsic, now
 the different rhythms of firings of the oscillators are only
effective and caused by  the different responses of the oscillators to pulses. Indeed a given pulse
causes  a larger phase advance on an oscillator with a smaller slope. Under the effect of pulses
oscillators with small slopes have larger effective rhythms than oscillators with large slopes.

Oscillators with close threshold values that avalanched together can remain locked in an avalanche
and form a stable group. The stability conditions for the whole system of $N$ oscillators are
similar to (\ref{conditiongroup}) and lead to the following probability $\cal{P}_N$ of complete
synchronization for a uniform distribution of slopes in $[a-\frac{D}{2},a+\frac{D}{2}]$:

\begin{eqnarray}
{\cal P}_N&=&\rho^{N-2}\int_0^{a_1}
ds_1\int_{s_1}^{2a_1}ds_2\dots\int_{s_{N-3}}^{(N-2)a_1}ds_{N-2} e^{-\rho s_{N-2}}\\
&=&1-e^{-\rho a_1}-\rho a_1 e^{-2\rho a_1}-\sum_{j=2}^{N-1}\frac{(j+1)^{j-1}}{j!}
(\rho a_1 e^{-\rho a_1})^je^{-\rho a_1},
\label{probaamplitude}
\end{eqnarray}
with $\rho={N\over D}$ and $a_1=a_{min}=a-\frac{D}{2}$. ${\cal P}_N$  depends on $D\over a$
through $\rho a_1=N({a\over D}-{1\over 2})$ and is independent of the dissipation parameter
$\alpha$. 
$P_N$ goes to $1$ with increasing $N$ and for a finite population size the model does not
synchronize only for very large disorder, typically $D\sim 2a$. 

\begin{figure}[t]
$$\includegraphics*[width=7cm]{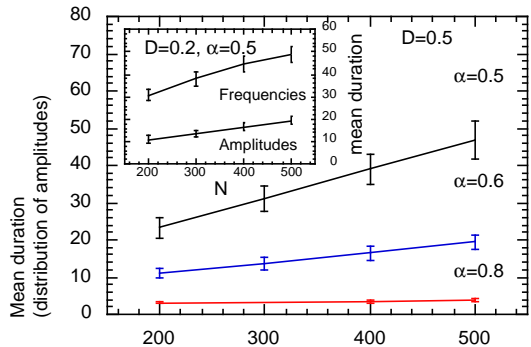}$$
\caption[DurŽes des transitoires, oscillateurs avec dŽsordre]{\renewcommand{\baselinestretch}{1}\sl Mean durations $T_S$ until synchronization for the model with a distribution of amplitudes of
width $l=0.5$ around the unit for {$\alpha=0.5,0.6,0.8$}(top to bottom) as a function of
the population size
$N$. Inset: mean durations to synchronization for the models with a distribution of amplitudes
(bottom) and of frequencies (top) for a distribution width $D=0.2$  and {$\alpha=0.5$}. For the
frequencies $T_S\simeq 19+0.06 N$ }
\label{DurationFreqandThresh}
\end{figure}

In short, we see that as in the model with a distribution of frequencies, we found that the duration
of the transient
$T_S$ until synchronization increases linearly with $N$  (fig.~\ref{DurationFreqandThresh}). For
identical $\alpha$ and  $D/a$, $T_S$  is shorter in the case with disorder on the amplitudes than on
the frequencies (fig.~\ref{DurationFreqandThresh} inset). $T_S$ depends strongly on the dissipation
$\alpha$. However we do not have  enough data for a precise relationship.

As in  the model with a distribution of frequencies, complete synchronization occurs
independently of the initial values of the phases (states) if the locking conditions of all the
oscillators in a single group are fulfilled.  The conditions for this locking depend however on the
models. Starting the evolution of the system from random phases, the formation of the possible stable
groups comes from the evolution of the relative phase gaps between the oscillators due to
different  rhythms which have their origin in the quenched disorder on the
characteristics of the oscillators. In the model with 
 different slopes they come
from the different phase advance responses to pulses of the oscillators.

For a given level of disorder and same $\alpha$ the probability of synchronization is much higher in
the case of a disorder on the amplitudes (thresholds) than on the frequencies with also shorter
$T_S$ (fig.\ref{DurationFreqandThresh} inset). In short, disorder on the amplitudes and slopes is
not a strong restriction of synchronization which is  much more limited by the disorder on the
frequencies. It is however not possible to conclude directly
on  what happens when both disorder exist simultaneously and  we shall now  therefore study this 
case.

\subsection{Disorder on the frequencies and  amplitudes}
\label{frequencythresholds}
The mechanism of synchronization that we saw at work in systems with two different kinds of disorder
is still at work and leads also to synchronization  in a system with mixed disorder as well on  the
frequencies as on the thresholds. 
For two oscillators $O_i$ and $O_j$ as previously, the first return map for the phase of $O_j$ is
now 
\begin{equation}
\phi_j^{k+1}=\phi_j^k+\Delta_{i,j}
\label{firstreturnmixed}
\end{equation}
with 
\begin{equation}
\Delta_{i,j}=(\phi_j^c-\phi_i^c)+\frac{a_j-a_i}{a_ia_j}\delta
\label{deltamixed}
\end{equation}
The phase variation $\Delta_{i,j}$ is due now as well to the difference of  the free frequencies
(first term of (\ref{deltamixed})) as to the different response of oscillators of different slopes to
pulses (second term of (\ref{deltamixed})). Since there is no relation between the signs of
$(\phi_j^c-\phi_i^c)$ and of  $({a_j-a_i})$ the two terms may be opposite. But generically they do
not cancel each other since the periods and slopes are random. The phase gap between $O_i$ and $O_j$
varies therefore monotonically and both oscillators avalanche at some time together.

\begin{figure}[t]
$$\includegraphics*[width=7cm]{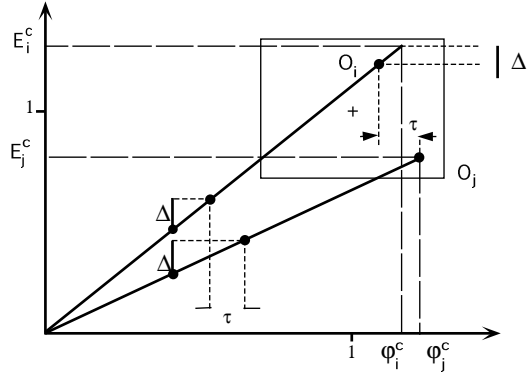}$$
\caption[Oscillateurs avec dŽsordre sur les frŽquences et les amplitudes]{\renewcommand{\baselinestretch}{1}\sl Oscillators with
 a spread of periods and thresholds
$(\phi_i^c,a_i)\in[1-l/2,1+l/2]^2$. Two oscillators $O_i$ and $O_j$ with
$\phi_i^c<\phi_j^c$ and $a_i>a_j$ have different locking conditions according to the strength of the
dephasing pulse $\Delta$. The locking condition depends on the order of the firings which depends on
$\Delta$. For a small$\Delta$, $O_i$ is at the threshold before $O_j$. For a large $\Delta$ (case 
represented in the figure), the oscillator $O_j$ which has the largest period is first at the
threshold.}
\label{figuremixeddisorder}
\end{figure}

\begin{figure}[t]
$$\includegraphics*[width=7cm]{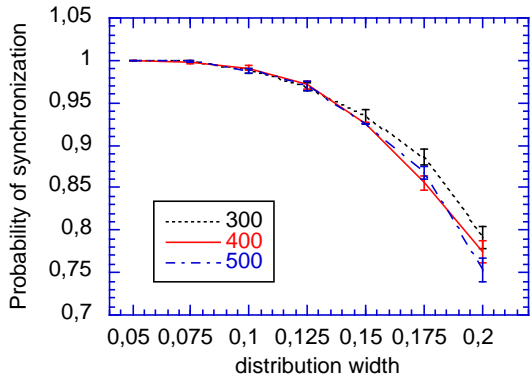}$$
\caption[ProbabilitŽ de synchronisation avec dŽsordre mixte]{\renewcommand{\baselinestretch}{1}\sl Probability of synchronization for a system of $N$ oscillators with a  uniform spread of
periods and thresholds $(\phi_i^c,a_i)\in[1-l/2,1+l/2]^2$, with $N=300,400$ as a
function of the width $l$ for $\alpha=0.5$. Each point is the probability obtained with
$1000$ samples of random oscillator parameter and is represented with the statistical error.}
\label{graphmixeddisorder}
\end{figure}

Here also there are locking conditions of oscillators in avalanches so that stable groups form and
may grow up to complete synchronization. However it is not possible in this case to get the 
probability of complete synchronization  proceeding as previously  by simply establishing the locking
conditions for all the $N$ oscillators in an avalanche. These conditions  are necessary but not
sufficient anymore to ensure synchronization for any initial distribution of the oscillator states.
Indeed  it is possible to verify that for large disorder there are cases were the formation of a
stable group between two oscillators $O_i$ and
$O_j$ with $a_i>a_j$ actually depends on the configuration of the phase values in the system and of
its history (fig.\ref{figuremixeddisorder}).
We studied the probability of synchronization numerically on simulations with
random thresholds and periods uniformly distributed in
$[1-D/2,1+D/2]*[1-D/2,1+D/2]$.  We see on fig.\ref{graphmixeddisorder} for $N=300,400,500$ and
$\alpha=0.5$ that up to
$D\sim 0.1$  complete synchronization occurs in more than $99\%$ of the cases and that the
probability is still high for larger widths. 
In the cases without complete synchronization the
asymptotic behavior consists in the periodic avalanches  of a large stable group with some few
small ones. For small disorder the probability of synchronization depends only weakly on $N$: 
for a given  $D$  the probabilities found for $N=300,400,500$ are all within the statistical errors.

\begin{figure}[t]
$$\includegraphics*[width=7cm]{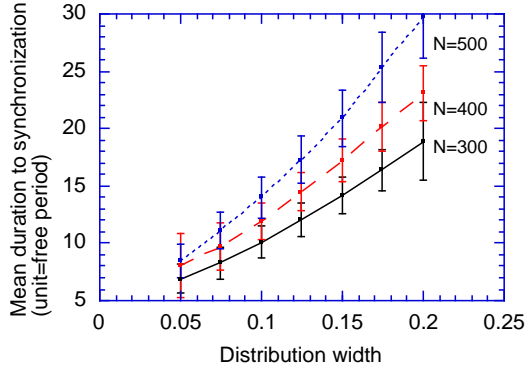}$$
\caption[DurŽes des transitoires, dŽsordre mixte]{\renewcommand{\baselinestretch}{1}\sl Mean duration $T_S$ until synchronization for the models with disorder on the frequencies and
on the thresholds $(\phi_i^c,a_i)\in[1-l/2,1+l/2]^2$ as a function of the
distribution width for $N=500,400,300$ and $\alpha=0.5$. $T_S$ grows polynomially with $D$, for
$N=500$, $T_S\simeq 4.5+63 D +316 D^2$ with correlation coefficient $0.99992$.}
\label{durationmixeddisorder}
\end{figure}

As seen on fig.~\ref{durationmixeddisorder} the duration of the transient occurs in only a few
periods although it  increases polynomially with the disorder width. While  the duration increases
also with the population size $N$, we do not have enough data to establish a precise relation.

At this point it is not difficult to imagine other models that
synchronize following the same principles. A model with a distribution of frequencies and slopes and
with constant threshold as been presented in
\cite{BottaniI}. We can also consider disorder on the pulse strengths. Let us, for instance, take a
population of identical oscillators with a quenched disorder only on the pulse strengths so that the
firing on an oscillators $O_i$ transmits to the rest of the system a pulse of strength
$\alpha_i/N$. The phase gap between two oscillators $O_i$ and $O_j$ varies then as
$s_{i,j}=(\alpha_i-\alpha_j)/N$. Since generically the $s_{i,j}\neq0$  any two oscillators
participate at some time in the same avalanche and since the slopes and thresholds are identical the
two oscillators are automatically locked.

We see finally that with instantaneous pulses, models with quenched disorder on several oscillator
characteristics may also evolve to synchronization by the same mechanism of evolution of the gaps and
locking in stable groups. The analysis and the estimation of the probability of synchronization is
however more complicated.

\section{Conclusion}
\label{Conclusion}
In this article we have highlighted on some simple models the existence of several mechanisms
leading to synchronization of IF oscillators.  A surprising result is that contrary to a common
belief, synchronization can actually occur even in basic models and  for identical oscillators 
independently of the shape of the oscillators. In particular oscillators 
do not need to have a convex evolution function in order to
synchronize\footnote{Synchronization with a general shape $E(\phi)$ has been also recently proven by
  Corral {\sl et al.}\cite{CorralI} in the case of oscillators with adequate
  response function to pulses. A response function chosen so that the phase
  advance caused by a pulse gets larger towards the threshold is equivalent to
  a convex $E(\phi)$.}. Therefore the common
interpretation that ``leakiness'' in the evolution of the free oscillators, which implies convexity,
is necessary for synchronization should be revised. We conclude that the observation of
synchronization in a system of IF oscillators implies by itself nothing about the shape of the
oscillator internal state variation function $E(\phi)$. Actually, for very concave oscillators  
synchronization occurs very easily  for the natural choice of initial random phases. It is the
opposite for a random initial distribution of the states. Therefore  the nature of the random 
configuration at the beginning of the evolution has possibly important consequences. It would be
interesting to study if
 the nature of the random initial configuration has similar consequences also in more sophisticated
models.

In this article we assumed direct additivity of the pulses, which is probably an excessive
requirement for realistic applications. The positive feedback effect between groups of different
sizes, which is the only mechanism of synchronization for linear oscillators, occurs also for a
softer form of additivity where the pulse from one group is not directly proportional to the number
of oscillators in the group but merely an increasing function of it. Softer additivty would however
increment the duration of the transient towards synchronization and reduce the range of favorable
parameters in the case with disorder. 

Concerning additivity we see that it is nevertheless true that convexity favors synchronization,
since it is the only case which synchronizes  also without additivity. However without additivity,
{\sl i.e.} without positive feedback, the duration  of the transient diverges then at least as
$(1-a)^{-1}$ in the linear limit
$a\rightarrow 1$ \cite{MirolloII}. Therefore without additivity a large
convexity is  necessary to keep the durations of the transitory not too long.

Let us mention that as shown recently by  Tsodyks {\sl et 
al.}\cite{Tsodyks},  Hansel {\sl et al.}\cite{Hansel}  and Abbott and van
Vreeswijk \cite{Abbott} smooth pulses with finite rise and fall time can crucially affect the
behavior and destabilize synchronization. In this article we assume fast interactions and  absorption: when two
oscillators fire one after the other, the pulse of the second one occurs entirely during the
refractory time of the first so that the oscillators synchronize in phase. The existence of a
refractory time and absorption ({\sl i.e.} assumption of fast pulses) is however not necessary 
for synchronization  for identical convex and linear oscillators, in which case synchronization
occurs also without absorption as locking of the oscillators in stable avalanches, in other words
this corresponds to out-of-phase locking of the oscillators. 

For identical linear and concave oscillators the probability of synchronization depends entirely on
the initial configuration of the phases/states of the system.  Indeed, some sets of initial
configurations  do not synchronize, for instance  when the initial phases are equally
spaced  so that no group can be  formed or in cases where the evolution leads to configurations
with groups of the same size. For linear and highly concave $E(\phi)$ the measure of the unfavorable
initial configuration is vanishingly small. It is larger and limits the probability of
complete synchronization for ``moderate'' concavity.
The degeneracy of the non favorable
configuration disappears if some disorder is included in the models such as a small spread on
the frequencies, thresholds or  pulse strengths.  

With fast pulse we found indeed that synchronization is possible also with a
range of disorder on the oscillator characteristics. 
 We find that the most difficult situation for synchronization is when the
oscillators have different frequencies, where, for small disorder, a system with a given random
sample of frequencies synchronizes almost always but, for larger disorder, the  probability of
synchronization decreases rapidly. Synchronization occurs then in the form of locking in avalanches
and should be affected by softer additivity. 
On the other hand, disorder on the shape of the oscillators  --
occurring here through disorder on the thresholds and hence different slopes--   with
otherwise identical frequencies does not constraint severely synchronization. When both kinds of
disorder are mixed the probability of complete synchronization is limited by the spread on the
frequencies.

A point of interest would be to investigate how the discussed effects occur in more complex realistic
models for instance of biological relevance. In particlar it would be important to study the
robustness of the results when the interaction pulses have finite rise and fall times and for
systems that are not globally coupled.

\medskip
ACKNOWLEDGMENTS:\\
I am grateful to B. Delamotte for his helpful advice and supervision. I am also indebted to 
C. P\'erez and A. D\'\i az-Guilera for fruitful discussions. Laboratoire de Physique Th\'eorique
et Hautes Energies is a Unit\'e associ\'ee au CNRS: D 0280.

\appendix 

\section{APPENDIX A: PROOF OF SYNCHRONIZATION FOR SMALL CONCAVITY}
\label{appendixconcave}
In this appendix we prove the synchronization of an assembly of $N$ identical oscillators with
state evolution function $E(t)=t^a, t\in[0,1], a>1$ in the limit $a\rightarrow 1$.
Let us first study the synchronization of only two
isolated groups
$G_i$ and
$G_j$ of respectively
$N_i$ and $N_j$ oscillators with $ (N_i>N_j)$  in absence of any other exterior pulses.
\begin{table}
{\parindent=-20pt
$$\begin{tabular}{|lcc|}
\hline
&$\mathbf{G_i}$&$\mathbf{G_j}$\\
\hline
{$G_i$ at threshold}&$1$&$\phi_j^{k}$\\
{Relaxation of $G_i$}&$0$&$\bigg[(\phi_j^{k})^a+N_i\delta\bigg]^{1\over a}$ \\
{$G_j$ at threshold}&$1-\bigg[(\phi_j^{k})^a+N_i\delta\bigg]^{1\over a}$&$1$\\
{Relaxation of $G_j$}&$\Big(1-\bigg[(\phi_j^{k})^a+N_i\delta\bigg]^{1\over a}+
N_j\delta\Big)^{1\over a}$&$0$\\
{$G_i$ at threshold}&$1$&$1-\Big(1-\bigg[(\phi_j^{k})^a+N_i\delta\bigg]^{1\over a}+
N_j\delta\Big)^{1\over a}$\\
\hline
\end{tabular}$$}
\caption[Application de premier retour, oscilateurs concaves]{\renewcommand{\baselinestretch}{1}\sl Evolution of the phases of two isolated groups $G_i$ and $G_j$ of $N_i$ and $N_j$
oscillators $(N_i>Nj)$  on a cycle where the firing of the first group does not succeed to drag the
second one along in an avalanche $((\phi_j^{k})^a+N_i\delta<E_c=1)$.}
\label{TableConcave}
\end{table}
In table \ref{TableConcave}, we trace the variation of the phases and  state variables of the two
groups on a cycle of relaxations beginning with the largest group at the threshold. We deduce from
there that the first return map for the phase of the second group $G_j$ is
\begin{equation}
\phi_j^{k+1}=1-\Big(1-\bigg[(\phi_j^k)^a+N_i\delta\bigg]^{1\over a}
+N_j\delta\Big)^{1\over a}. 
\label{FirstReturnConcave}
\end{equation}
which has an attractive fixed point $\phi_0(a,N_i,N_j,\delta)$.
If the new phase after a cycle $\phi_j^{k+1}$ is in the interval $I_c\equiv[\phi_c(a,N_i,\delta),1]$
where
$\phi_c(a,N_i,\delta)=(1-N_i\delta)^{1\over a}$ then $G_j$ is absorbed in the relaxation of $G_i$
(see fig.\ref{fixedpointconcave}).
\begin{figure}[t]
\mbox{\includegraphics[width=6cm]{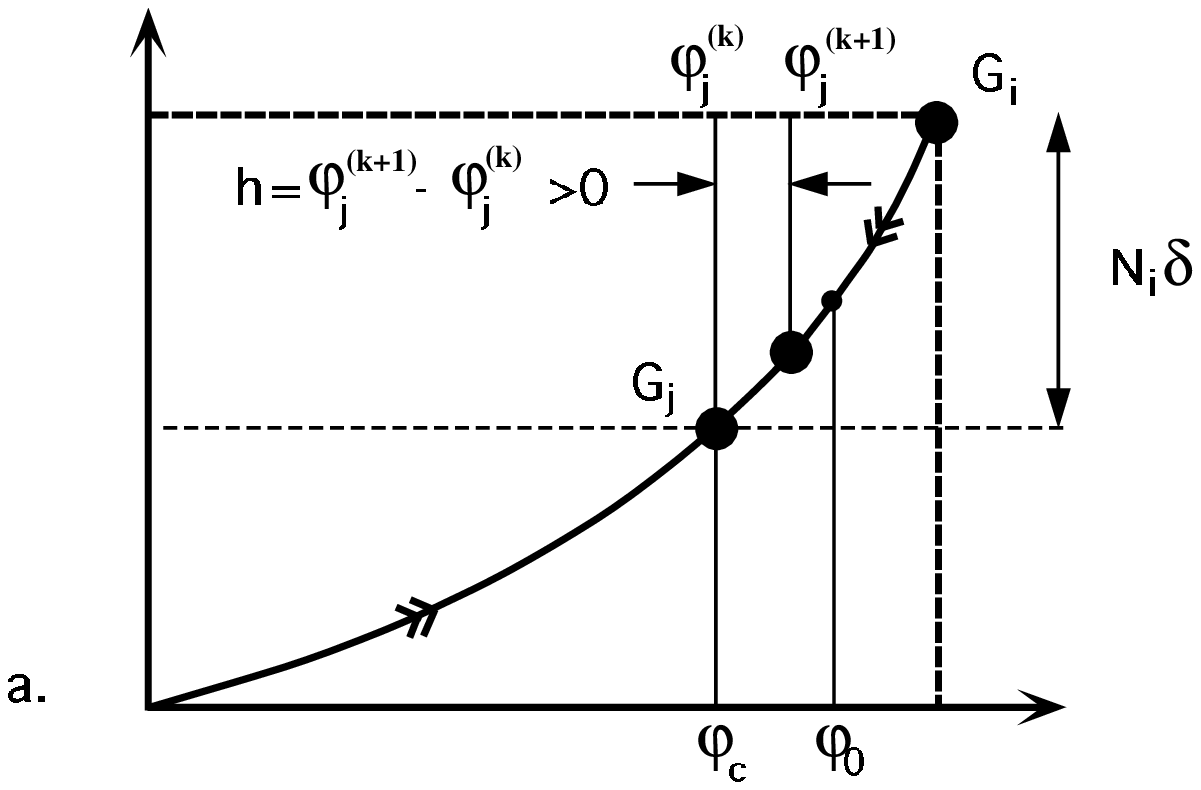}}\mbox{\includegraphics*[width=6cm]{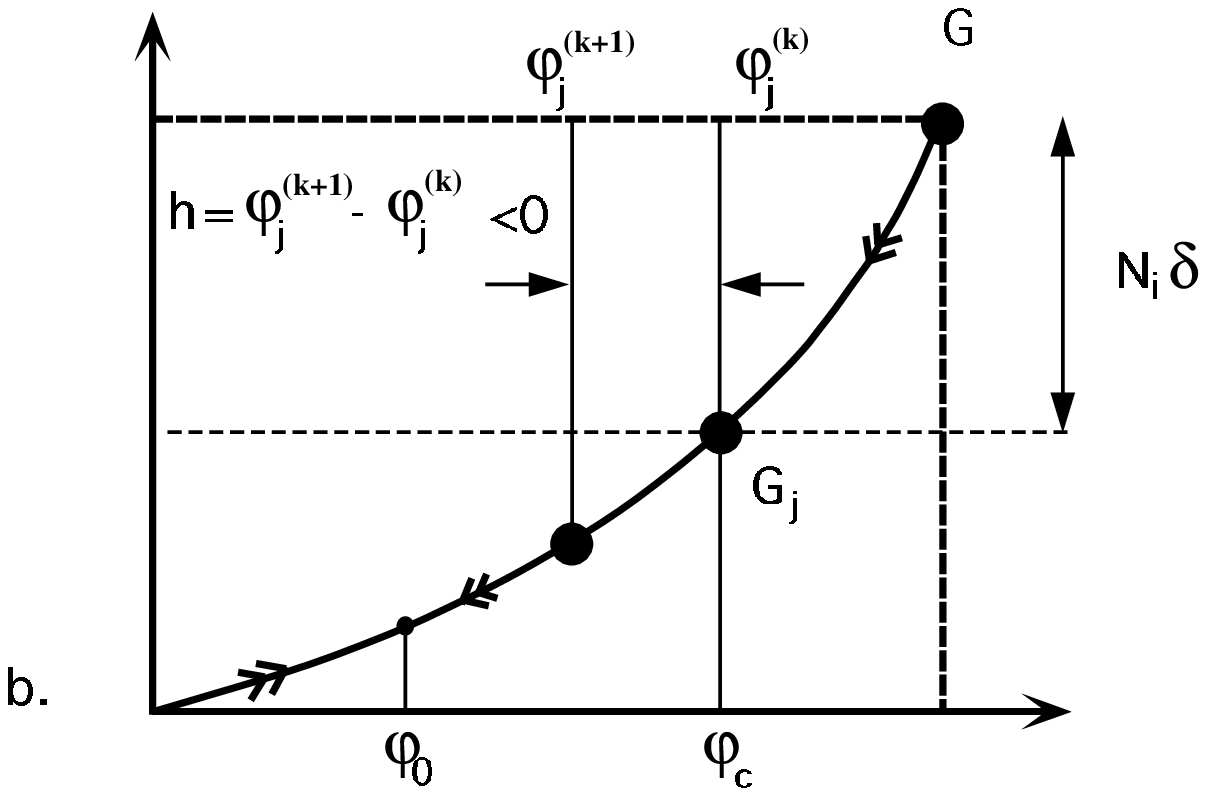}}
\caption[Oscillateurs avec dŽsordre sur les frŽquences et les amplitudes]{\renewcommand{\baselinestretch}{1}\sl Test for the synchronization of two groups for concave oscillators. {\bf a.} The firing of
$G_1$ causes the avalanche of the second group $G_2$ if $E_2+N_1\delta \ge 1$, that is if
$\phi_j^{(k+1)} \ge \phi_c$. $\phi_0$ is the fixed point of the first return map for the phase of $G_2$ on each cycle
beginning with $G_1$ at the threshold. {\bf b.)} If $\phi_j^{(k+1)}<\phi_c$ the second groups never
comes sufficiently close to the first one in order to avalanche together.d in the figure), the oscillator $O_j$ which has the largest period is first at the
threshold.}
\label{fixedpointconcave}
\end{figure} 

$\phi_c$ corresponds to the phase at which $G_j$ is just pushed at the threshold by the pulse of
$G_i$. If
$\phi_0>\phi_c$, then  the gap between the two groups gets smaller on each repeated cycle
until it becomes sufficiently small for the groups to avalanche together and to merge. On the other
hand if $\phi_0<\phi_c$ the two groups never avalanche together and remain apart.
It is analytically difficult to test directly if $\phi_0$ is in $I_c$. However,  since
(\ref{FirstReturnConcave}) is monotonic on each side of the fixed point, it is more convenient
to test
 if
$\phi_j^{k+1}$ is in $I_c$ when $\phi_j^k=\phi^c$, i.e.  is just at the border of $I_c$.  With
$\phi_j^k=\phi_c$ this gives the inequality:
\begin{equation}
h(N_i,N_j)\equiv \phi_j^{k+1}-\phi_j^k=1-(1-N_i\delta)^{1\over a}-(N_j\delta)^{1\over a} \ge 0,
\label{TestSynchronyConcave}
\end{equation}
where $h(N_i,N_j)$ is the variation of the  phase of $O_j$ on a cycle assuming $\phi_j^k=\phi_c$.
Let us examine (\ref{TestSynchronyConcave}) for two groups  with sizes $(N+c)/2$
and $(N-c)/2$. The function $g(c)\equiv  h((N+c)/2,(N-c)/2)$ is monotonically increasing in $c$ so
that the attraction between the two groups is stronger when the size difference is bigger. 
For a given  $N$ and $a$ the condition $g(c)\ge 0$ is fulfilled when $c\ge \bar{c}(a,N)$ with
\begin{equation}
\bar{c}=\frac{N(1-a)}{2\alpha a}
\left((1-\frac{\alpha}{2})\ln\left(1-\frac{\alpha}{2}\right)+\frac{\alpha}{2}
\ln\left(\frac{\alpha}{2}\right)\right)
+o\left((a-1)^2\right).
\label{cbar}
\end{equation}
Contrary to the case of linear oscillators, two groups of different sizes -- not only of equal size --
may remain apart and not synchronize: if their size difference is too small,  i.e. 
$c<\bar{c}$,  positive feedback is not efficient enough and absorption does not occur. 
$\bar{c}(a,N)$ is an increasing function of $a$, so that for larger concavities less configurations
with two groups can synchronize. We will see that this determines the probability that a system with
$N$ random initial phases synchronizes.

Up to now we have considered only two isolated groups.
In order to see if synchronization can occur when there are more groups, let us choose two groups 
and see if at some time they merge together.  With many groups it is not possible to write a simple
first return function on a cycle for the gap between  two successive groups  since this return map
depends sensitively on the history of the system during this cycle. However we can simplify the
question and prove that two groups can merge by focusing on the most severe condition. For
that, let us isolate the two groups from the rest of the system as if they would not be affected by
the pulses from the oscillators outside of the pair.
It is easy to see that if the two groups can
synchronize in these circumstances, they still synchronize in the real situation with the influence
of exterior pulses.  Indeed, the pulses of the rest of the system increment all the states in the
same way and so they do not change the gap $E_i(t)-E_j(t)$.  Therefore if $O_i$ and
$O_j$ are close enough to avalanche   together, they do so independently of pulses of other
oscillators in the system. We  can therefore focus our study on the case of an
isolated pair of oscillators. 
 Let the sizes of the two groups be $n$ and $n-c$. The groups merge if
(\ref{concavlimit}) is fulfilled with $N_i=n$ and $N_j=n-c$. Differently from previously we
now study (\ref{TestSynchronyConcave}) with $N_i+N_j\neq N$. Since $h(n,n-c)$ is again a monotonically
increasing function of $c$, the bigger is $c$, the stronger is the attraction. Therefore
the most stringent condition for synchronization is for two groups of minimal size difference.
Keeping this in mind we should now examine (\ref{TestSynchronyConcave}) as a function of $n$, i.e. we
examine $f(n)\ge 0$ with $f(n)\equiv h(n,n-c)$ when $n\in [c+1,(N+c)/2]$. The function $f(n)$ is
monotonically decreasing  on the variation interval of $n$ with  the highest value $f(c+1)\ge 0$.
The smallest value $f(\frac{N+c}{2})$ is equal to $g(c)$. This value is the change in phase  that we
studied previously for a system of  two groups  of sizes $\frac{N+c}{2}$ and $\frac{N-c}{2}$. If the
condition $g(c)\ge 0$ is fulfilled, then $f$ is also positive over the whole interval $n\in
[c+1,(N+c)/2]$ and any pair of groups with size difference $c$ synchronizes. 
Finally we see that it is for the case of only two groups of sizes $\frac{N+c}{2}$ and $\frac{N-c}{2}$
that synchronization is the most difficult and it is this case that determines the most stringent
condition for this phenomenon.
Therefore, assuming that, as in the case of linear oscillators, synchronized pairs spontaneously form 
during the first cycle we find that the probability that the system
synchronizes completely for  random initial  phases corresponds to  the probability that
$c<\bar{c}(a,N)$. Unfortunately this is also difficult to
calculate.
The system synchronizes with highest probability if synchronization is possible even for two groups
of sizes $(N+1)/2$ and $(N-1)/2$, that is if $g(c=1)\ge 0$. This is the case when
\begin{equation}
a<\bar{a}\equiv
1-\frac{\alpha}{N}\left(\left((1-\frac{\alpha}{2}\right)\ln\left(1-\frac{\alpha}{2}\right)+
\frac{\alpha}{2}\ln\left(\frac{\alpha}{2}\right)\right)^{-1}+o\left(\alpha\frac{1}{N^2}\right).
\label{concavlimit}
\end{equation}
Since $\bar{a}>1$ there is  an interval of concavities with the same 
conditions of synchronization than the linear case. Then synchronization can stop only  if the
two last groups are of  the same size. Since
$\bar{a}$ is close to
$1$, the corresponding  range of concave functions is quite small. However as discussed in
\ref{Concave},  synchronization  occurs in practice also with high probability for much larger
concavities.

\section{APPENDIX B: DISTRIBUTION OF AMPLITUDES}
\label{appendixamplitudes}
In this appendix we detail the conditions under which synchronization occurs in the model of section
(\ref{amplitudes}) of oscillators with a distribution of amplitudes (thresholds).
We follow the same steps  as  for the model with a distribution
of frequencies (\ref{frequencies}).
\begin{table}[t]
$$\begin{tabular}{|cccr|}
\hline
&$\mathbf{\phi_i}$&$\mathbf{\phi_j}$&\\
\hline
{\bf a.}&$1$&$\phi_j^{(k)}$&\\
{\bf b.}&$0$&$\phi_j^{(k)}+\delta/a_j$&\\
{\bf c.}&$1-\phi_i^{(k)}+(j-2)\delta\frac{a_j-a_i}{a_ja_i}-\frac{\delta}{a_j}$&$1$&\\
{\bf d.}&$1-\phi_i^{(k)}+(j-2)\delta\frac{a_j-a_i}{a_ja_i}+
\frac{\delta}{a_i}-\frac{\delta}{a_j}$&$0$&\\
{\bf d.}&$1$&$\phi_j^{(k)}+(N-1)\delta\frac{a_j-a_i}{a_ja_i}$&\\
\hline
\end{tabular}$$
\caption[Application de premier retour, desordre d'amplitudes]{\renewcommand{\baselinestretch}{1}\sl {\bf a.} $O_i$ at the threshold; {\bf b.} Firing of $O_i$ ,  $O_j$ received a pulse $\delta$
causing an advance in phase $\delta/a_j$; {\bf c.} $O_j$ at the threshold, the term
$(j-2)\delta\frac{a_j-a_i}{a_ja_i}$ comes from the $(j-2)$ other pulses since the relaxation of
$O_i$; {\bf d.}  Firing of  $O_j$; {\bf e.} $O_i$ back at the threshold;  $(N-j+1)$ other pulses from
the rest of the system  occurred since the relaxation of $O_i$.}
\label{TableAmpl}
\end{table}
 From table \ref{TableAmpl} we get 
the first return map for the phase of $O_j$ on a cycle beginning with $O_i$ at the threshold:
\begin{equation}
\phi_j^{(k+1)}=\phi_j^{(k)}+(N-1)\delta\frac{a_j-a_i}{a_ja_i}.
\label{firstreturnampl}
\end{equation}
Let $a_j>a_i$, then on each cycle $\phi_j$ is
closer to
$\phi^c=1$. The phase difference $\phi_i-\phi_j$ decreases and
 after some repetitions of the cycle  the firing of $O_i$ drags $O_j$  along in an avalanche. 
Therefore as in section \ref{frequencies} also in this model any two oscillators avalanche at some
time together. The change in the phase gaps between the oscillators that finally cause the
simultaneous firings  has for origin the different  rhythms of firings of the oscillators. Contrary
to the previous model where the different rhythms were intrinsic, now
 the different rhythms of firings of the oscillators are only
effective and caused by  the different responses of the oscillators to pulses.
Indeed the value of 
the phase advance caused by a pulse of given strength  depends on the slopes of the oscillators. 
Due to the quenched disorder on the slopes  the oscillator evolve more or less rapidly under the
phase advance caused by   pulses and have therefore different
effective rhythms of evolution.   The evolution towards synchronization due to the different rhythms
comes in addition to the positive feedback attraction between groups of different sizes  which
causes also the evolution of the  phase gaps between  oscillators. Both effects drive the system
in the state of maximal synchronization compatible with the disorder. 

We establish  now   the stability conditions of synchronized groups, i.e. the conditions of locking
in avalanches. Two oscillators $O_i$ and $O_j$, say with $a_j>a_i$ that avalanche  together and are in
phase at the origin
$\phi_i=\phi_j=0$   are  dephased by the pulses from other oscillators,  the oscillator $O_i$ with
the smallest slope being the most advanced.  Let $\Delta$ be the summed strength of the pulses of the
other oscillators between the last simultaneous avalanche of $O_i$ and $O_j$ and the return of $O_i$
back at the threshold.
$\Delta$ shifts the two oscillators apart by the phase difference $\tau=\Delta(a_i-a_j)/(a_ia_j)$.
If the  slopes $a_i$ and $a_j$ are close enough then $\tau$ is sufficiently small
for $O_i$ and $O_j$  still to relax in the same avalanche  triggered by $O_i$. The locking condition
for two oscillators that avalanched  together is (see fig.
\ref{distribampl}):
\begin{equation}
\Delta\frac{a_i-a_j}{a_ia_j}<\frac{\delta}{a_j}.
\label{condpairampl}
\end{equation}
If $O_i$ and $O_j$ were the only oscillators in their avalanche then $\Delta=(N-2)\delta$ and
(\ref{condpairampl}) is equivalent to ${a_i-a_j} < a_i/(N-2)$.

For a group of  $m\ge 2$ oscillators $O_i,i=1\dots
m$ with $a_{i+1}>a_i$ the locking conditions are:
\begin{equation}
(a_i-a_1)<a_1\frac{i-1}{N-m}\ \ \ , \ \ \ i=1\dots m.
\label{condgroupampl}
\end{equation}
These inequalities are obtained considering that  \begin{enumerate}
\item the $m$ oscillators that avalanched together and
were at the origin are dephased by a total pulse $\Delta=(N-m)\delta$  before the
oscillators $O_1$  is back, the first of them, at the threshold.
\item the $i$-th oscillator in the avalanche receives $i-1$ pulses from the oscillators that
preceded it.
\end{enumerate}
Complete synchronization is possible if (\ref{condgroupampl}) is fulfilled for $m=N-1$. 
Indeed, if
this is the case a stable group with $N-1$ elements  forms. Then, this group and the last $N$-th
oscillator of the system inevitably participate in a same avalanche and the whole system becomes in
phase without, now, any exterior dephasing  pulse. 

The relation (\ref{probaamplitude}) in section \ref{amplitudes}  gives the probability for a uniform
random distribution
 of $N$ slopes in an interval $[a-\frac{D}{2},a+\frac{D}{2}]$ of fulfilling (\ref{condgroupampl}).


\end{document}